\documentclass[10pt,3p,times]{elsarticle}

\usepackage{amsmath,amsfonts,amssymb}
\usepackage{amsthm}
\usepackage{bm}
\usepackage[toc]{appendix}
\usepackage[T1]{fontenc}
\usepackage{textcomp}
\usepackage{graphicx,epstopdf}
\usepackage[position=top,labelfont=normalfont,textfont=normalfont,singlelinecheck=off,justification=raggedright]{subfig}
\usepackage{floatrow}
\usepackage[dvipsnames]{xcolor}
\usepackage{setspace}
\usepackage{enumerate,etaremune}
\usepackage{tabularx,multirow}
\usepackage{framed}
\usepackage{hhline}
\usepackage[unicode,bookmarks=false]{hyperref}
\usepackage{url}
\hypersetup{
  colorlinks,
  citecolor=Blue,linkcolor=Blue,urlcolor=Blue
}
\usepackage{lineno}

\usepackage{tikz}
\usetikzlibrary{shapes.geometric}
\usetikzlibrary{shapes,arrows}
\usetikzlibrary{plotmarks}

\usetikzlibrary{calc}

\usepackage{algorithm}
\usepackage{algorithmicx,algpseudocode}
\usepackage{cases}

\newcommand{\eg}{{\it e.g.}}
\newcommand{\ie}{{\it i.e.}}
\newcommand{\cf}{{\it cf.}}
\newcommand{\etal}{{\it et al.}}
\newcommand{\tensor}[1]{\bm{#1}}
\newcommand{\stress}{\sigma}
\newcommand{\strain}{\varepsilon}
\newcommand{\tstress}{\tensor{\stress}}
\newcommand{\tstrain}{\tensor{\strain}}

\newcommand{\od}{\mathrm{d}}
\newcommand{\pd}{\partial}

\newcommand{\el}{\mathrm{e}}

\newcommand{\romannumber}[1]{\uppercase\expandafter{\romannumeral #1\relax}}
\newcommand{\cn}{\mathrm{N}}

\newcommand{\bulk}{\mathrm{bulk}}
\newcommand{\inter}{\mathrm{interface}}
\newcommand{\fric}{\mathrm{friction}}
\newcommand{\nopene}{\mathrm{no-penetration}}
\newcommand{\slip}{\mathrm{slip}}

\newcommand{\sfe}{\mathcal{G}_{\romannumber{2}}}
\newcommand{\yield}{\mathrm{Y}}

\DeclareMathOperator{\grad}{\nabla}
\DeclareMathOperator{\diver}{\grad\cdot}
\DeclareMathOperator{\symgrad}{\nabla^{s}}

\DeclareMathOperator{\dyad}{\otimes}
\newsavebox{\dotbox}

\theoremstyle{remark}
\newtheorem{remark}{Remark}

\newcommand{\revised}[1]{{\color{black} #1}}

\newcolumntype{L}[1]{>{\raggedright\let\newline\\arraybackslash\hspace{0pt}}m{#1}}
\newcolumntype{C}[1]{>{\centering\let\newline\\arraybackslash\hspace{0pt}}m{#1}}
\newcolumntype{R}[1]{>{\raggedleft\let\newline\\arraybackslash\hspace{0pt}}m{#1}}

\allowdisplaybreaks

\biboptions{sort&compress,square,comma,numbers}
\AtBeginDocument{\hypersetup{citecolor=MidnightBlue,linkcolor=MidnightBlue,urlcolor=MidnightBlue}}

\usepackage{etoolbox}
\makeatletter
\patchcmd{\ps@pprintTitle}
{Preprint submitted to}
{~}
{}{}
\makeatother

\begin{document}

\begin{frontmatter}

\title{A phase-field model of frictional shear fracture in geologic materials}

\author[HKU]{Fan Fei}
\author[HKU]{Jinhyun Choo\corref{corr}}
\ead{jchoo@hku.hk}

\cortext[corr]{Corresponding Author}

\address[HKU]{Department of Civil Engineering, The University of Hong Kong, Hong Kong}

\journal{~}

\begin{abstract}
Geologic shear fractures such as faults and slip surfaces involve marked friction along the discontinuities as they are subjected to significant confining pressures.
This friction plays a critical role in the growth of these shear fractures, as revealed by the fracture mechanics theory of Palmer and Rice decades ago.
In this paper, we develop a novel phase-field model of shear fracture in pressure-sensitive geomaterials, honoring the role of friction in the fracture propagation mechanism.
Building on a recently proposed phase-field method for frictional interfaces, we formulate a set of governing equations for different contact conditions (or lack thereof) in which frictional energy dissipation emerges in the crack driving force during slip.
We then derive the degradation function and the threshold fracture energy of the phase-field model such that the stress--strain behavior is insensitive to the length parameter for phase-field regularization.
This derivation procedure extends a methodology used in recent phase-field models of cohesive tensile fracture to shear fracture in frictional materials in which peak and residual strengths coexist and evolve by confining pressure.
The resulting phase-field formulation is demonstrably consistent with the theory of Palmer and Rice.
Numerical examples showcase that the proposed phase-field model is a physically sound and numerically efficient method for simulating shear fracture processes in geologic materials, such as faulting and slip surface growth.
\end{abstract}

\begin{keyword}
Phase-field models \sep
Shear fracture \sep
Frictional contact \sep
Geologic materials \sep
Faults \sep
Slip surfaces
\end{keyword}

\end{frontmatter}


\section{Introduction}
The initiation and growth of geologic shear fractures (\eg~faults and slip surfaces) are a common trigger of various subsurface failures including onshore and offshore landslides~\cite{Skempton1964,Bjerrum1967,Cooper1998,Veveakis2007,Locat2011,Quinn2011,Dey2015,Borja2016b}.
They are also central to a number of geomorphological processes, having significant implications in structural geology and other relevant fields~\cite{Segall1980,Paterson2005,Pollard2005,Sanz2007,Sanz2008,Schultz2019}.
Predictive modeling of shear fracture processes in geologic materials is thus an important task for scientists and engineers alike.

Shear fractures in geologic materials display a few important characteristics that distinguish them from fractures considered by classic linear elastic fracture mechanics theory or cohesive zone models.
First, the two surfaces of a geologic shear fracture are usually in contact with friction---unlike those of an opening fracture---because geologic materials in the field are subjected to some amount of confining pressure.
This frictional contact not only gives rise to the residual shear strength of a crack but also plays a critical role in the mechanics of fracture propagation.
Second, unlike most other types of fractures that only nucleate under tensile stress, shear fractures in geologic materials can develop under compressive stress as a consequence of crack initiation, growth, and coalescence from preexisting flaws~\cite{Menendez1996,Wu2000,Healy2006,Wong2009c,Lee2011}.
It is noted that indeed the majority of geologic shear fracture processes take place under significant compressive loads.
Third, the peak and residual shear strengths of a geomaterial depend strongly on the current confining pressure because the major portion of the strengths stem from the frictional resistance between grains.
Due to these distinct characteristics, dedicated efforts are required to accurately model shear fractures in geologic materials.

\begin{figure}[htbp]
    \centering
    \includegraphics[width=1.0\textwidth]{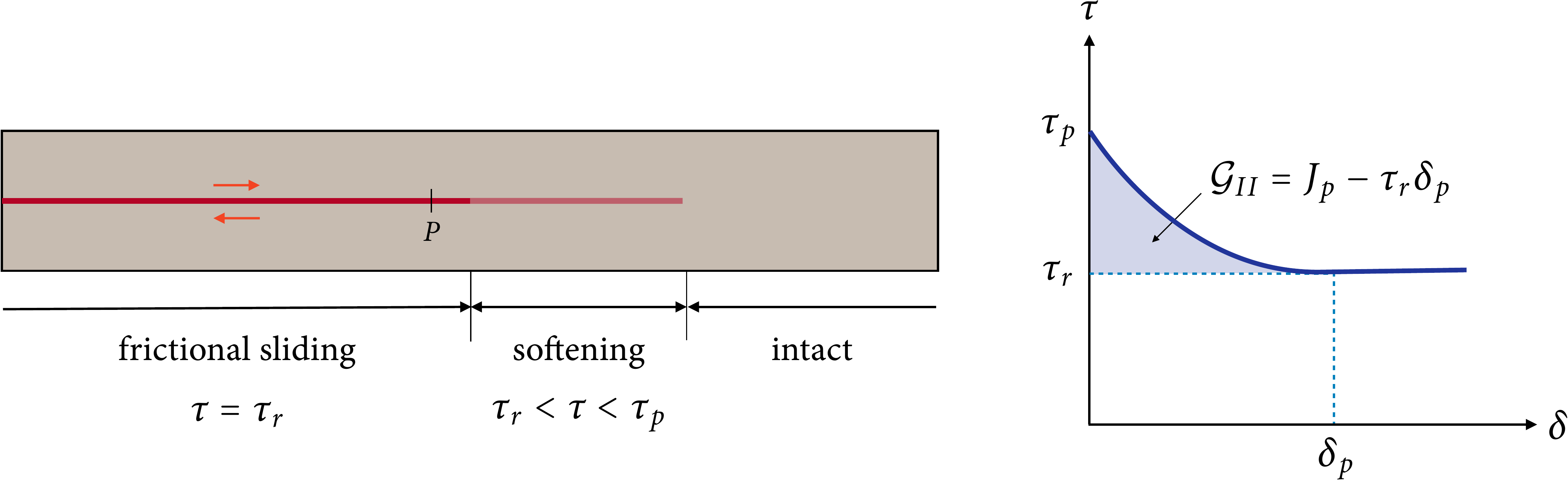}
    \caption{Schematic illustration of Palmer and Rice's theory~\cite{Palmer1973}.
    (Left) During shear fracturing, frictional sliding takes place along an existing slip surface while the process zone undergoes strain softening.
    The shear stress ($\tau$) equals the residual shear strength ($\tau_{r}$) along the slip surface, and it ranges between the residual shear strength and the peak shear strength ($\tau_{p}$) in the process zone.
    (Right) The shaded area in the shear stress versus slip ($\delta$) curve corresponds to the shear fracture energy ($\sfe$) of the material.
    The shear fracture energy is equal to the energy release rate at Point $P$ ($J_{p}$) minus $\tau_{r}\delta_{p}$ where $\delta_{p}$ is the slip at Point $P$.}
    \label{fig:palmer-rice-illustration}
\end{figure}

In 1973, Palmer and Rice~\cite{Palmer1973} formulated a fracture mechanics theory for the growth of slip surfaces in overconsolidated clays.\footnote{Technically speaking, Palmer and Rice~\cite{Palmer1973} referred to a slip surface as a shear band. However, in the modern geomechanics literature, the term shear band is usually reserved for a tabular zone of localized strain, which is distinguished from a sharp discontinuity such as a slip surface (see~\cite{Borja2004,Aydin2006,Schultz2008} for example). For this reason, the shear discontinuity of our interest is called a shear fracture in this paper.
Note, however, that it is not uncommon in geomechanics to treat a shear band as a sharp (strong) discontinuity, because the thickness of a shear band---which is in the order of grain size---is often negligible for field-scale problems. If this viewpoint is adopted, the phase-field model developed in this work may also be applied to shear bands in brittle geomaterials.}
Using the $J$-integral approach~\cite{Rice1968}, they derived an energy-based condition that governs the propagation of a preexisting slip surface.
The crux of their theory is the explicit incorporation of the frictional energy during slip into the propagation condition, which is absent in other theories for tensile fractures for an obvious reason.
They have formally shown that the energy release rate during the growth of a slip surface is the sum of the shear fracture energy and the frictional energy, and that the shear fracture energy can be estimated from a shear stress--slip curve as illustrated Fig.~\ref{fig:palmer-rice-illustration}.
The theory of Palmer and Rice has been well applied to a number of problems related to the propagation of slip surfaces and faults (\eg~\cite{McClung1979,Wong1982a,Abercrombie2005,Rice2006,Viesca2012}).
Furthermore, since early 2000s, Puzrin and coworkers have generalized the theory to more ductile geomaterials~\cite{Puzrin2005} and applied it to pioneer a new method for geomechanical stability analyses which they termed the shear band propagation approach (\eg~\cite{Puzrin2004,Puzrin2011,Puzrin2016,Puzrin2016b,Puzrin2017,Puzrin2019}).
The approach has shown remarkable success in analyzing a wide variety of geologic hazard problems, which proves the physical soundness of the underlying theory.

Nevertheless, the fracture mechanics theory of Palmer and Rice~\cite{Palmer1973} has not been well incorporated into computational models of shear fractures in geologic materials.
The vast majority of models in the literature has relied on Mohr--Coulomb theory for shear strength and/or its plasticity versions to simulate the development of a frictional crack (\eg~\cite{Regueiro1999,Borja2001,Wu2012,Wu2013}).
Although Mohr--Coulomb theory can reliably capture the pressure-dependent nature of the peak shear strength, it does not inform one how strain-softening takes place in the fracture process zone.
So Mohr--Coulomb theory alone is insufficient to model the process of shear fracture propagation.
Borja and coworkers~\cite{Borja2007,Foster2007,Liu2009} addressed this issue by inserting a slip weakening law from Palmer and Rice's theory into elements that undergo softening after bifurcation.
It is however noted that this approach still deviates from the theory of Palmer and Rice~\cite{Palmer1973}, because the propagation condition is not based on energy dissipation.
Also, such post-bifurcation enrichment requires the use of a discrete numerical method for discontinuities such as the assumed enhanced strain method~\cite{Simo1993,Regueiro2001,Borja2002} or the extended finite element method~\cite{Dolbow2001,Liu2008,Liu2010,Sanborn2011,Liu2013,Prevost2016}, plus explicit treatment of frictional contact along a discontinuity.
Unfortunately, in this kind of discrete method, algorithms for enriching basis functions and tracking discontinuous geometry are often onerous to implement and possible sources of bias.
The algorithmic difficulty becomes amplified when the discontinuity evolves with frictional contact.

In recent years, phase-field models of fracture have emerged as a continuous method for simulating the formation and growth of a crack without an algorithm to track the crack geometry (\eg~\cite{Bourdin2008,Miehe2010,Borden2012}).
Many phase-field models have also been developed for geomaterials under mechanical and non-mechanical loads (\eg~\cite{Lee2016b,Santillan2017,Choo2018b,Ha2018}), but they have focused only on tensile fractures.
A few notable exceptions are as follows.
Zhang \etal~\cite{Zhang2017} and Bryant and Sun~\cite{Bryant2018} proposed phase-field models of mixed-mode (tensile and shear) fractures and applied them to simulate the growth of cracks in laboratory-scale specimens.
While these models successfully reproduced shear and mixed-mode fractures as observed from experiments, they neglected the pressure dependence of geomaterials.
Choo and Sun~\cite{Choo2018a} introduced a framework that couples phase-field modeling with pressure-sensitive plasticity, demonstrating that it can simulate an array of failure modes of geomaterials---from tensile fracture to shear fracture to ductile flow---and their transition by confining pressure.
Yet the coupled phase-field and plasticity framework did not explicitly distinguish between shear and tensile fractures, although they are known to be very different for geomaterials.

Critically, the existing phase-field formulations for shear fractures in geologic materials are not consistent with the fracture mechanics theory of Palmer and Rice~\cite{Palmer1973}.
The main reason is that none of them has explicitly and accurately incorporated friction along a fracture and its contribution to the fracture propagation process.
Only very recently, Fei and Choo~\cite{Fei2020} have proposed a stress-decomposition scheme for phase-field modeling of frictional cracks and interfaces.
However, the stress-decomposition scheme focused mainly on modeling frictional contact in a given interface, rather than the propagation of a frictional crack.
Indeed, any combination of the stress-decomposition scheme with an existing  phase-field model will still be inconsistent with the theory of Palmer and Rice~\cite{Palmer1973}, because its stress--strain response must be sensitive to the length parameter introduced for phase-field regularization.
In other words, the peak and residual shear strengths of such combination will inevitably depend on the length parameter introduced for a diffuse approximation of sharp geometry.
However, these strengths should be independent of the length parameter because they are prescribed material parameters in the Palmer and Rice's theory (\cf~Fig.~\ref{fig:palmer-rice-illustration}).

In this work, we develop the first phase-field formulation for shear fracture in pressure-sensitive geomaterials that is fully consistent with the fracture mechanics theory of Palmer and Rice~\cite{Palmer1973}.
The development entails two unprecedented, formidable challenges in phase-field modeling of fracture: (i) incorporation of the slip-induced frictional energy into the fracture propagation mechanism, and (ii) length-insensitive modeling of the stress--strain response in the presence of peak and residual strengths that depend on confining pressure.
To address the first challenge, we build on the stress-decomposition scheme for diffuse interfaces with frictional contact~\cite{Fei2020}, and derive a set of governing equations for different contact conditions (or lack thereof) with an explicit consideration of frictional energy under a slip condition.
Subsequently, to tackle the second challenge, we extend a methodology originally developed for cohesive tensile fracture that render a phase-field formulation virtually insensitive to the phase-field length parameter (\eg~\cite{Geelen2019,Wu2017,Wu2018}), which themselves are built on cohesive-type gradient damage formulations pioneered by Lorentz and coworkers~\cite{Lorentz2011,Lorentz2011a,Lorentz2012,Lorentz2017}.
Note that the extension of the existing methodology is a remarkable task in that here we must deal with two strengths---peak and residual shear strengths---that evolve by confining pressure during the course of loading, instead of a single, constant peak strength as in cohesive tensile fracture.
It will be shown that the resulting phase-field formulation is remarkably consistent with Palmer and Rice's theory.

The paper is organized as follows.
In Section~\ref{sec:formulation}, we develop the formulation of the phase-field model, describing the main contributions of this work.
In Section~\ref{sec:discretization}, we discretize the formulation and devise algorithms for numerical implementation of the proposed model.
In Section~\ref{sec:examples}, we present numerical examples to verify the phase-field model and investigate its capabilities for simulating shear fracture processes in geologic materials under various conditions.
We conclude the work in Section~\ref{sec:closure}.

\section{Phase-field Formulation}
\label{sec:formulation}
The purpose of this section is to develop a new phase-field formulation for shear fracture in geologic materials under confining pressure.
Particular attention is paid to achieving full consistency with the fracture mechanics theory of Palmer and Rice~\cite{Palmer1973}.
Therefore, as in the theory, we will consider quasi-brittle geomaterials in which plasticity and finite deformations are negligible.
Inertial effects will also be ignored.

\subsection{Phase-field approximation}
Consider a body $\Omega\in\mathbb{R}^{\mathrm{dim}}$ where dim denotes the spatial dimension.
The external boundary of the body, $\partial\Omega$, is suitably decomposed into the displacement (Dirichlet) boundary, $\partial_{u}\Omega$, and the traction (Neumann) boundary, $\partial_{t}\Omega$, such that $\overline{\partial_{u}\Omega\cap\partial_{t}\Omega}=\emptyset$ and $\overline{\partial_{u}\Omega\cup\partial_{t}\Omega}=\partial\Omega$.
The body may have internal discontinuities whose set is denoted by $\Gamma$.
Let us denote by $\mathbb{T}:=[0,t_{\max}]$ the time domain of interest and $t\in\mathbb{T}$ the current time.

Following the standard phase-field modeling approach, we diffusely approximate the sharp geometry of $\Gamma$ by introducing a phase-field variable, $d\in[0,1]$.
The phase-field variable denotes a completely discontinuous (interface) region by $d=1$ and a completely continuous (bulk) region by $d=0$.
Then we introduce a surface density functional $\Gamma_{d}(d,\grad d)$, whose general form can be written as
\begin{align}
  \Gamma_{d}(d,\grad d) := \dfrac{1}{c_0}\left[\dfrac{\alpha(d)}{L} + L\grad{d}\cdot\grad{d}\right]\,, \quad \text{where}\;\;
  c_{0} := 4\int_{0}^{1}\sqrt{\alpha(s)}\,\od s\,.
  \label{eq:crack-density-functional-general}
\end{align}
Here, $L$ is the length parameter controlling the width of the phase-field approximation zone,
$c_0$ is a normalization constant,
and $\alpha(d)$ is the local dissipated energy density.
The local dissipation term is chosen as $\alpha(d) := d^2$ in standard phase-field models of brittle fracture (\eg~\cite{Bourdin2008,Miehe2010,Borden2012}).
It is now well-known that this choice of $\alpha(d)$ makes the peak strength of the material sensitive to the phase-field length parameter, $L$~\cite{Borden2016,Mandal2019}.
However, such length-sensitivity is undesirable for phase-field modeling of cohesive fracture because the peak strength of a cohesive zone model is a prescribed material parameter which should not be affected by $L$.
This is also the case in this work because we aim at modeling shear fracture with prescribed peak and residual shear strengths.
Therefore, here we use $\alpha(d) := d$, which was proposed by Bourdin \etal~\cite{Bourdin2014} and then applied by Geelen \etal~\cite{Geelen2019} for length-insensitive phase-field modeling of dynamic cohesive fracture.
Inserting $\alpha(d) := d$ into Eq.~\eqref{eq:crack-density-functional-general}, we get the following specific form of the crack density function:
\begin{align}
  \Gamma_{d}(d,\grad d) = \dfrac{3}{8}\left[\dfrac{d}{L} + L\grad{d}\cdot\grad{d}\right]\,. \quad
  \label{eq:crack-density-functional-specific}
\end{align}
It is noted that other forms of $\alpha(d)$ have also been advanced for length-insensitive phase-field modeling of cohesive tensile fracture.
Notable examples are presented in the work of Wu~\cite{Wu2017}.
Here we have chosen $\alpha(d) := d$ for its relative simplicity, and we expect that other forms of $\alpha(d)$ may be equally well employed in the succeeding model development as long as they are compatible with length-insensitive phase-field modeling.

\subsection{Stress decomposition in frictional interfaces}
For phase-field modeling of shear fracture under compressive stress, we must model the frictional behavior of a diffusely approximated interface.
For this purpose, we adopt the stress-decomposition scheme proposed by Fei and Choo~\cite{Fei2020}, which is a simple yet accurate method for incorporating frictional contact into phase-field modeling.
In what follows, we describe the essence of the stress-decomposition scheme, noting some points that specialize in a strain-softening material depicted in Fig.~\ref{fig:palmer-rice-illustration}.

In phase-field modeling of fracture, the overall stress tensor at a material point, $\tstress$, is calculated by interpolating the stress tensor in the bulk (intact) region, $\tstress_{\bulk}$, and the stress tensor in the interface (fully damaged) region, $\tstress_{\inter}$.
This interpolation can be written as
\begin{align}
    \tstress = g(d)\tstress_{\bulk} + [1 - g(d)]\tstress_{\inter}\,,
    \label{eq:stress-interpolation}
\end{align}
where $g(d)$ is the so-called degradation function subjected to the following constraints:
\begin{align}
    g(d)\in[0,1]\,, \quad
    g(0) = 1\,, \quad
    g(1) = 0\,, \quad
    g'(d) \leq 0\,.
    \label{eq:degradation-function-generic}
\end{align}
One can see that any form of $g(d)$ satisfying the above conditions gives $\tstress = \tstress_{\bulk}$ in an intact region where $d=0$,
and $\tstress=\tstress_{\inter}$ in a fully damaged region where $d=1$.
The most common form of $g(d)$ in the phase-field literature is $g(d)=(1-d)^2$.
However, this form of $g(d)$ inevitably makes the stress--strain response sensitive to the phase-field length parameter.
Indeed, to suppress the length-sensitivity of a phase-field formulation, the form of $g(d)$ should be carefully derived~\cite{Geelen2019,Wu2017,Wu2018}.
Later in this section, we will select a generic form of $g(d)$ compatible with length-insensitive modeling and derive its specific expression for our phase-field formulation for shear fracture.

The bulk stress tensor is calculated using a standard stress--strain constitutive model in continuum mechanics.
\revised{A general stress--strain relationship can be written in the rate form as}
\begin{align}
    \dot{\tstress}_{\bulk} = \mathbb{C}_{\bulk}:\dot{\tstrain}\,,
\end{align}
where $\mathbb{C}_{\bulk}$ is the fourth-order stress--strain tangent tensor and $\tstrain$ is the infinitesimal strain tensor defined as the symmetric gradient of the displacement vector, $\tensor{u}$.
\revised{For simplicity, we assume that the pre-failure behavior is well described by isotropic linear elasticity.}
Then $\mathbb{C}_{\bulk}$ equals the linear elastic stiffness tensor, $\mathbb{C}^{\el}$, given by
\begin{align}
    \mathbb{C}_{\bulk} = \mathbb{C}^{\el} := K\tensor{1}\dyad\tensor{1} + 2G\left(\mathbb{I} - \dfrac{1}{3}\tensor{1}\dyad\tensor{1} \right)\,,
\end{align}
where $K$ and $G$ are the bulk modulus and the shear modulus, respectively, $\tensor{1}$ is the second-order identity tensor, and $\mathbb{I}$ is the fourth-order symmetric identity tensor.
The bulk and shear moduli can be converted into other elasticity parameters such as Young's modulus $E$ and Poisson's ratio $\nu$.
We note that the assumption of isotropic linear elasticity can be relaxed if one wants to capture more realistic pre-failre behavior~\cite{Choo2011,Choo2013}.

\begin{figure}[htbp]
  \centering
  \includegraphics[width=0.85\textwidth]{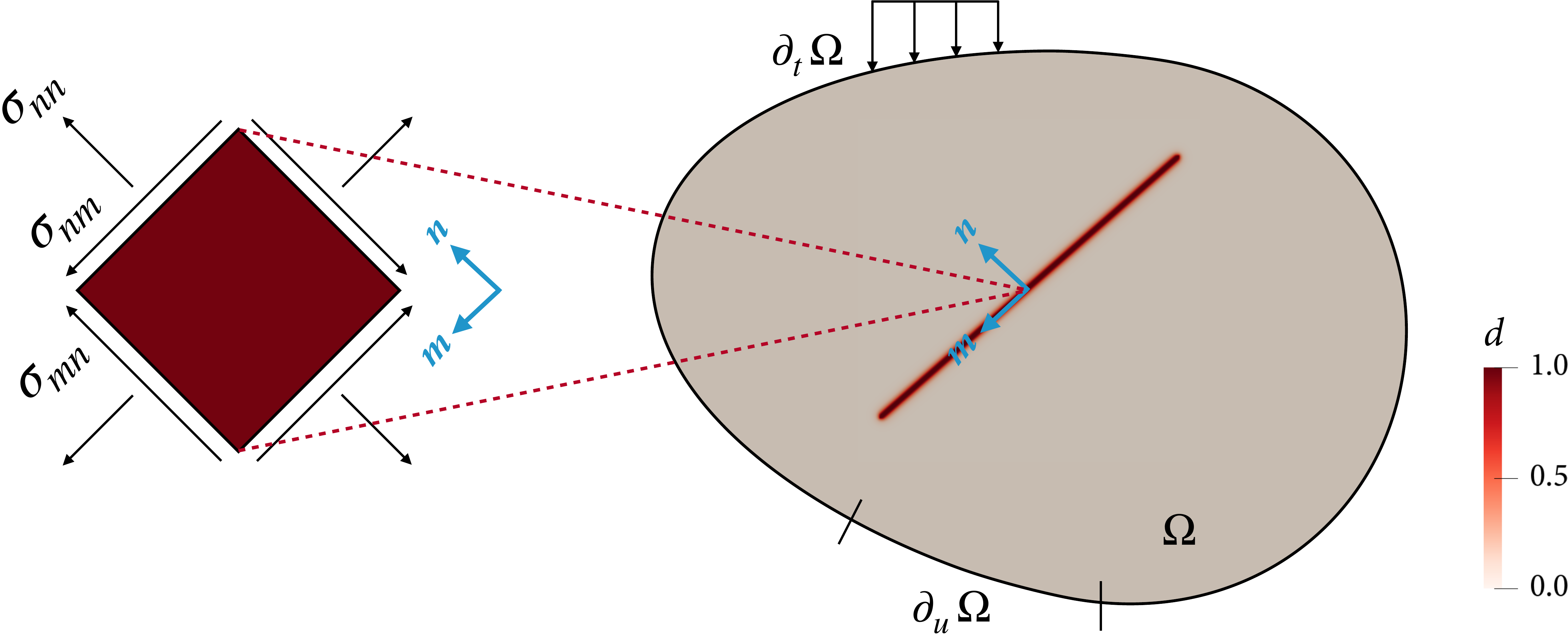}
  \caption{An interface-oriented coordinate system in a domain where discontinuous geometry is diffusely approximated by the phase-field variable, $d$. Vectors $\tensor{n}$ and $\tensor{m}$ denote unit vectors in the interface normal and tangential/slip directions, respectively. Also shown are the definitions of some stress components in the interface-oriented coordinate system. After Fei and Choo~\cite{Fei2020}.}
  \label{fig:interface-coord}
\end{figure}

The interface stress tensor is determined according to the contact condition of the interface.
To identify the contact condition, we introduce a new coordinate system oriented to the normal and tangential directions of the interface, which we refer to as an interface-oriented coordinate system.
Figure~\ref{fig:interface-coord} illustrates an example of an interface-oriented coordinate system along with the unit normal vector, $\tensor{n}$, and the slip vector, $\tensor{m}$, of an interface.
The upshot of introducing an interface-oriented coordinate system is that we can compute the stress and strain components in the contact normal and tangential directions.
Then we can define the contact normal strain, $\strain_{\cn}$, as
\begin{align}
    \strain_{\cn} \equiv \strain_{nn} = \tstrain:(\tensor{n}\dyad\tensor{n})\,.
\end{align}
Using $\strain_{\cn}$ as the gap function in classic contact mechanics, we consider the interface is open when $\strain_{\cn} > 0$ and in contact otherwise.
When the interface is in contact, we further distinguish between stick and slip conditions using a yield function that gives $f<0$ under a stick condition and $f=0$ under a slip condition.
For a strain-softening material depicted in Fig.~\ref{fig:palmer-rice-illustration}, the yield function can be written as
\begin{align}
    f := |\tau| - \tau_{\yield} \leq 0\,.
\end{align}
Here, $\tau$ is the overall resolved shear stress, calculated as
\begin{align}
    \tau := \frac{1}{2}\tstress:\tensor{\alpha}\,,\quad \text{where}\;\;
    \tensor{\alpha} := \tensor{m}\dyad\tensor{n} + \tensor{n}\dyad\tensor{m}\,,
\end{align}
and $\tau_{\yield}$ is the yield stress.

The interface stress tensor is then calculated depending on the contact condition as follows:
\begin{align}
    \tstress_{\inter}
    = \left\{\begin{array}{ll}
        \tensor{0} & \text{if} \;\; \text{open}\,, \\ [0.5em]
        \tstress_{\bulk} & \text{if} \;\; \text{stick}\,, \\ [0.5em]
        \tstress_{\fric} + \tstress_{\nopene} & \text{if} \;\; \text{slip}\,.
    \end{array}\right.
\end{align}
The stress calculations in the first two cases (open and stick) are obvious.
In the last case (slip), the interface stress tensor is decomposed into the frictional component in the contact tangential direction, $\tstress_{\fric}$, and the no-penetration component in the contact normal direction, $\tstress_{\nopene}$.
To write specific expressions for the two components, we define
\begin{align}
    \tau_{\bulk} := \dfrac{1}{2}\tstress_{\bulk}:\tensor{\alpha}\,.
\end{align}
Then the frictional and no-penetration components are expressed as
\begin{align}
    \tstress_{\fric} = \tau_{r}\tensor{\alpha}\,, \quad
    \tstress_{\nopene} = \tstress_{\bulk} - \tau_{\bulk}\tensor{\alpha}\,,
\end{align}
because $\tau_{r}$ is the shear stress mobilized along a discontinuous slip surface.
Inserting these expressions into Eq.~\eqref{eq:stress-interpolation}, we get the expressions for the overall stress tensor as
\begin{align}
    \tstress
    = \left\{\begin{array}{ll}
        g(d)\tstress_{\bulk} & \text{if} \;\; \text{open}\,, \\ [0.5em]
        \tstress_{\bulk} & \text{if} \;\; \text{stick}\,, \\ [0.5em]
        \tstress_{\bulk} - [1 - g(d)]\tau_{\bulk}\tensor{\alpha} + [1 - g(d)]\tau_{r}\tensor{\alpha} & \text{if} \;\; \text{slip}\,.
    \end{array}\right.
    \label{eq:overall-stress-tensor}
\end{align}
It is noted that for the expression under a slip condition, the first two terms are related to the bulk deformation whereas the last term is related to the slip in the interface.
Also, from the last expression in Eq.~\eqref{eq:overall-stress-tensor}, we can write the overall resolved shear stress, $\tau$, as
\begin{align}
    \tau = g(d)\tau_{\bulk} + [1 - g(d)]\tau_{r}\,,
    \label{eq:tau-1d}
\end{align}
which is consistent with the interpolation for the overall stress tensor, Eq.~\eqref{eq:stress-interpolation}.
One can see that $\tau=\tau_{\bulk}$ in an intact zone where $d=0$, $\tau_{r}< \tau < \tau_{p}$ in the process zone where $0<d<1$ and $\tau_{\bulk}>\tau_{r}$, and $\tau=\tau_{r}$ in a fully developed slip surface where $d=1$.
This is also consistent with the shear stress distribution in the problem of slip surface growth in a strain-softening material, depicted in Fig.~\ref{fig:palmer-rice-illustration}.
\smallskip

\revised{
\begin{remark}
    The foregoing treatment of frictional contact assumes that dilation is absent in frictional slip. This assumption is standard in contact mechanics and also consistent with the theory of Palmer and Rice~\cite{Palmer1973}. However, fractures in geologic materials often show some amount of shear-induced dilation due to surface roughness or gouge materials.
    The effect of fracture dilation would be small, but it can be critical in some cases such as coupled poromechanical problems.
    Extension of the current formulation to dilatant fractures will be presented in future work.
\end{remark}
}

\subsection{Governing equations}
Following the work of da Silva \etal~\cite{daSilva2013}, we will obtain the two governing equations of the phase-field model from the balance of macroscopic momentum as
\begin{align}
    \diver\,\left(\dfrac{\pd{\psi}(\tstrain,d,\grad{d})}{\pd\tstrain}\right) - \dfrac{\pd\psi(\tstrain,d,\grad{d})}{\pd\tensor{u}} = \tensor{0}\,,
    \label{eq:macro-momentum-balance}
\end{align}
and from the balance of microscopic force as
\begin{align}
    \diver\,\left(\dfrac{\pd{\psi}(\tstrain,d,\grad{d})}{\pd\grad{d}}\right) - \dfrac{\pd\psi(\tstrain,d,\grad{d})}{\pd d}
    = \left\{\begin{array}{ll}
      \beta\dot{d} & \text{if} \;\; \dot{d}>0\,,\\
      -\pi_{r} & \text{if} \;\; \dot{d}=0\,.
    \end{array}
    \right.
    \label{eq:micro-force-balance}
\end{align}
Here, $\beta$ is a parameter related to the rate-dependence of the fracturing process, and we set $\beta=0$ in this work to consider rate-independent fracture.
Also, $\pi_{r}$ is the reactive microforce introduced for enforcing the crack irreversibility constraint.
Lastly, $\psi$ denotes the potential energy density at a material point.
For the problem at hand, it is given by
\begin{align}
    \psi = \psi^{\el} + \psi^{\mathrm{f}} + \psi^{\mathrm{d}} - \psi^{\mathrm{b}},
    \label{eq:potential-energy-density}
\end{align}
where $\psi^{\el}$ is the strain energy due to elastic deformation, $\psi^{\mathrm{f}}$ is the frictional energy due to slip, $\psi^{\mathrm{d}}$ is the fracture energy due to the generation of discontinuous surfaces, and $\psi^{\mathrm{b}}$ is the external energy due to body force.

Our task is then to determine specific expressions for the four energy terms in Eq.~\eqref{eq:potential-energy-density}.
Note that the first two energy terms, $\psi^{\el}$ and $\psi^{\mathrm{f}}$, have different forms according to the contact condition.
So we consider their rate forms to accommodate their incremental nonlinearity.
The rate of the strain energy density, $\psi^{\el}$, is given by
\begin{align}
    \dot{\psi}^{\el}
    = \left\{\begin{array}{ll}
        g(d)\tstress_{\bulk}:\dot{\tstrain} & \text{if} \;\; \text{open}\,, \\ [0.5em]
        \tstress_{\bulk}:\dot{\tstrain} & \text{if} \;\; \text{stick}\,, \\ [0.5em]
        \tstress_{\bulk}:\dot{\tstrain} - [1 - g(d)]\tau_{\bulk}\dot{\gamma}\ & \text{if} \;\; \text{slip}\,,
    \end{array}\right.
    \label{eq:strain-energy-rate}
\end{align}
where $\gamma := \tstrain:\tensor{\alpha}$.
Next, the rate of the frictional energy density, $\dot{\psi}^{\mathrm{f}}$, can be written as
\begin{align}
    \dot{\psi}^{\mathrm{f}}
    = \left\{\begin{array}{ll}
        0 & \text{if} \;\; \text{open}\,, \\ [0.5em]
        0 & \text{if} \;\; \text{stick}\,, \\ [0.5em]
        [1 - g(d)]\tau_{r}\dot{\gamma} & \text{if} \;\; \text{slip}\,.
    \end{array}\right.
    \label{eq:frictional-energy-rate}
\end{align}
It is noted that this expression is a continuum version of the frictional energy considered in Palmer and Rice~\cite{Palmer1973}, for an interface region where $d=1$.
The last two energy terms, $\psi^{\mathrm{d}}$ and $\psi^{\mathrm{b}}$, are independent of contact conditions.
The fracture energy density is given by
\begin{align}
    \psi^{\mathrm{d}}
    = \sfe\Gamma_{d}(d,\grad{d})
    = \dfrac{3\sfe}{8L}[d + L^{2}\grad{d}\cdot\grad{d}]\,,
\end{align}
where $\sfe$ is the shear fracture energy.
Note that $\sfe$ depends heavily on the amount of slip, varying by orders of magnitude (see, Abercrombie and Rice~\cite{Abercrombie2005}, for example).
Within the same order of magnitude, it can also be changed by the normal stress and temperature~\cite{Wong1986,Zhang1990}.
In this work, for simplicity, we will prescribe $\sfe$ based on the expected amount of slip, neglecting its dependence on the normal stress and temperature.
Lastly, the external energy is given by
\begin{align}
    \psi^{\mathrm{b}} = \rho\tensor{g}\cdot\tensor{u}\,,
\end{align}
where $\rho$ is the mass density of the body, and $\tensor{g}$ is the gravitational acceleration vector.

Having determined the four energy terms, we now derive specific expressions for the governing equations~\eqref{eq:macro-momentum-balance} and~\eqref{eq:micro-force-balance}.
Due to the incrementally nonlinear nature of the post-peak behavior, the derivation must consider a sequence of loading/unloading.
In what follows, we will first consider a monotonic loading sequence in which a material point is intact ($d=0$) initially, undergoes slip-induced fracturing ($\dot{d}>0$) after shear stress reaches the peak shear strength, and becomes fully fractured ($d=1$) manifesting the residual shear strength.
Then we will consider an unloading sequence in which fracture does not grow ($\dot{d}=0$).
Note that the loading and unloading cases after the peak stress can be distinguished by the yield function, \ie~$f=0$ for loading (slip) and $f<0$ for unloading (stick), similar to classic plasticity.
For brevity, in what follows we will focus on a material point under compressive stress.
It is noted that fracturing under tensile stress can be addressed by an existing phase-field model of opening fracture.

\paragraph{Intact condition}
Consider an intact ($d=0$) region of which the contact condition cannot be defined due to the absence of an interface.
Assume that shear stress does not reach the peak shear strength in any direction.
For this intact region, we adopt the form of the governing equations of existing phase-field models of cohesive fracture (\eg~\cite{Wu2017,Wu2018,Geelen2019}) and write them as
\begin{align}
    \left.\begin{array}{rl}
        \diver\,\tstress_{\bulk} + \rho\tensor{g} = \tensor{0} \\ [0.5em]
        -g'(d)\mathcal{H}_{t} + \dfrac{3\sfe}{8L}\left(2L^2\diver \grad{d} - 1\right) = 0
    \end{array}\right\}
    \;\text{if}\;\; \text{intact}\,.
    \label{eq:governing-intact}
\end{align}
Here, $\mathcal{H}_{t}$ is the threshold fracture energy per unit volume, which is defined as the crack-driving part of the strain energy at the peak stress.
This term is introduced in phase-field models of cohesive fracture (\eg~\cite{Wu2017,Wu2018,Geelen2019}) to prevent fracturing until stress reaches a prescribed peak strength.
Note that shear fracture in Palmer and Rice's theory~\cite{Palmer1973} also nucleates only when shear stress reaches the peak shear strength.
While $\mathcal{H}_{t}$ for opening fracture can be straightforwardly determined as the tensile strain energy at the peak strength, its determination for shear fracture is not straightforward because not all part of the shear strain energy contributes to shear fracturing in which frictional energy dissipation exists.
Therefore, the specific form of $\mathcal{H}_{t}$ will be derived later when we furnish the formulation to be insensitive to the phase-field length parameter, $L$.
It is also noted that the same set of governing equations can be used for both loading and unloading in an intact region.

\paragraph{Slip (loading) condition}
Next, consider the subsequent stage of loading in which shear stress reaches the peak shear strength and then decreases to the residual shear strength.
During this strain-softening process, an interface region ($0<d\leq 1$) emerges and undergoes slip throughout.
To derive the governing equations in this process, we first have to determine the strain and frictional energies, $\psi^{\el}$ and $\psi^{\mathrm{f}}$.
To accommodate the pressure-induced evolution of $\tau_{r}$ during slip, we integrate the rate forms of the energy terms during the time of slip and add them to the energies when slip started, \ie~when shear stress in an intact region reached $\tau_{p}$.
Denoting by $t_{p}$ the time at the beginning of slip, the strain and frictional energies can be written as
\begin{align}
    \psi^{\el} &= \psi^{\el}|_{t_{p}} + \int_{t_p}^{t} \tstress_{\bulk}:\dot{\tstrain}\, \od t - \int_{t_{p}}^{t}\,[1 - g(d)]\tau_{\bulk}\dot{\gamma}\, \od t\,, \label{eq:strain-energy-slip} \\
    \psi^{\mathrm{f}} &= \psi^{\mathrm{f}}|_{t_{p}} + \int_{t_{p}}^{t}\,[1 - g(d)]\tau_{r}\dot{\gamma}\,\od t\,,
    \label{eq:frictional-energy-slip}
\end{align}
where $\psi^{\el}|_{t_{p}}$ and $\psi^{\mathrm{f}}|_{t_{p}}$ refer to the strain and frictional energies at $t_{p}$.
Obviously, $\psi^{\mathrm{f}}|_{t_{p}}=0$ because $\psi^{\mathrm{f}}=0$ in the beginning and $\dot{\psi}^{\mathrm{f}}=0$ before slip.
In addition, from the definition of $\mathcal{H}_{t}$ and Eq.~\eqref{eq:micro-force-balance}, we get
\begin{align}
    \dfrac{\pd \psi^{\el}|_{t_{p}}}{\pd d} = g'(d)\mathcal{H}_{t}\,.
    \label{eq:peak-strain-energy}
\end{align}
We then insert Eqs.~\eqref{eq:strain-energy-slip}--\eqref{eq:peak-strain-energy} into Eqs.~\eqref{eq:macro-momentum-balance} and~\eqref{eq:micro-force-balance}, and substitute $\dot{\gamma}=\od \gamma/\od t$ into the time-integrated equations.
The governing equations for post-peak loading ($f=0$) are obtained as
\begin{align}
    \left.\begin{array}{rl}
        \diver\,\{\tstress_{\bulk} - [1 - g(d)](\tau_{\bulk} - \tau_{r})\tensor{\alpha}\} + \rho\tensor{g} = \tensor{0} \\ [0.5em]
        -g'(d)\mathcal{H} + \dfrac{3\sfe}{8L}\left(2L^2\diver \grad{d} - 1\right) = 0
    \end{array}\right\}
    \;\text{if}\;\; \text{slip}\,.
    \label{eq:governing-slip}
\end{align}
where
\begin{align}
    \mathcal{H} := \mathcal{H}_{t} + \mathcal{H}_{\slip}\,, \quad
    \mathcal{H}_{\slip} := \int_{\gamma_{p}}^{\gamma} (\tau_{\bulk} - \tau_{r})\,\od \gamma\,.
    \label{eq:crack-driving-force}
\end{align}
Here, $\mathcal{H}_{\slip}$ is the crack driving force emanating from the slip motion, and $\mathcal{H}$ is the total crack driving force which also includes the crack driving force accumulated from the pre-peak deformation.
Also, $\gamma_{p}$ denotes the shear strain in the slip direction when $\tau=\tau_{p}$, such that the amount of frictional slip can be calculated as $\gamma - \gamma_{p}$.

\paragraph{Stick (unloading) condition}
Lastly, consider the case of post-peak unloading in which shear stress decreases after fracture has developed.
In this case, the interface is stick and the material point behaves as an intact material.
So we obtain $\psi^{\el} = \psi_{\bulk}$ and $\psi^{\mathrm{f}}=0$.
Also, because $\dot{d}=0$ during unloading, $\pi_{r}$ appears on the right hand side of the microforce equation~\eqref{eq:micro-force-balance} to ensure crack irreversibility.
Therefore, the governing equations in this case are given by
\begin{align}
    \left.\begin{array}{rl}
        \diver\,\tstress_{\bulk} + \rho\tensor{g} = \tensor{0} \\ [0.5em]
        \dfrac{3\sfe}{8L}\left(2L^2\diver \grad{d} - 1\right) = -\pi_{r}
    \end{array}\right\}
    \;\text{if}\;\; \text{stick}\,.
    \label{eq:governing-stick}
\end{align}
Following the algorithm of Miehe \etal~\cite{Miehe2010b}, which has widely been used in phase-field models of fracture, we set $\pi_{r}$ as
\begin{align}
    \pi_{r} = -g'(d) \max_{t\in[0,t]} \mathcal{H}(t)\,.
\end{align}
It is noted that the governing equations in this unloading case are equivalent to those in most existing phase-field models under compressive stress, in which the stress tensor is not degraded and the crack driving force is not updated.
However, in the current model, these equations arise only when the interface is stick.
\smallskip

\begin{remark}
    The crack driving force during slip, $\mathcal{H}_{\slip}$ in Eq.~\eqref{eq:crack-driving-force}, is expanded as
    \begin{align}
        \mathcal{H}_{\slip} = \int_{\gamma_{p}}^{\gamma}\tau_{\bulk}\,\od \gamma
        - \underbrace{\int_{\gamma_{p}}^{\gamma} \tau_{r}\,\od \gamma}_{\substack{\text{frictional}\\\text{energy}}}\,.
    \end{align}
    As noted in the above equation, the second term on the right hand side excludes the frictional energy dissipation from the crack driving force.
    Not only is this term conceptually consistent with Palmer and Rice's theory~\cite{Palmer1973}, but it also is quantitatively consistent because the frictional energy in the discrete theory is $\tau_{r}\delta$ (\cf~Fig.~\ref{fig:palmer-rice-illustration}) and $\gamma$ is the continuum counterpart of $\delta$.
    Note that this term has naturally emerged as a consequence of using the stress-decomposition scheme of Fei and Choo~\cite{Fei2020}.
\end{remark}

\subsection{Derivation of degradation function and threshold fracture energy}
\label{subsec:derivation-gd-Ht}
To complete the formulation, we further derive specific expressions for the degradation function, $g(d)$, and the threshold fracture energy, $\mathcal{H}_{t}$.
The derivation will be guided by the fact that standard continuum mechanics theory, which is adopted in this work as well as Palmer and Rice~\cite{Palmer1973}, does not have a material length scale.
Such absence of a material length scale requires that the stress--strain response of the phase-field formulation should not be sensitive to the phase-field length parameter, $L$.
In what follows, we show that this length-insensitivity argument allows us to arrive at specific expressions for $g(d)$ and $\mathcal{H}_{t}$.

We begin the derivation by adopting a generic form of $g(d)$ compatible with the crack density function~\eqref{eq:crack-density-functional-specific}, given by
\begin{align}
    g(d) = \dfrac{(1 - d)^n}{(1 - d)^n + md(1 + pd)}\,.
    \label{eq:degradation-function-form}
\end{align}
In this paper, we focus on two particular forms that were used by Geelen \etal~\cite{Geelen2019} for phase-field modeling of cohesive opening fracture and referred to as quasi-quadratic and quasi-linear functions.
The quasi-quadratic function is recovered by setting $n=2$ and $p\geq1$, and the quasi-linear one by $n=1$ and $p=0$.
For the quasi-quadratic function, $p$ acts as a material parameter controlling the softening behavior, \revised{which can be calibrated to fit experimental data. (See Geelen \etal~\cite{Geelen2019} for example.)}
The quasi-quadratic degradation function was originally proposed by Lorentz and coworkers~\cite{Lorentz2011,Lorentz2017}, while the quasi-linear degradation function was derived by Geelen \etal~\cite{Geelen2019}.
The remaining parameter, $m$, is not a free parameter and determined from other material properties.
To wit, consider an intact domain where $d=0$ throughout.
To satisfy the microforce balance in Eq.~\eqref{eq:governing-intact}, it is required that
\begin{align}
    m = \dfrac{3\sfe}{8L}\dfrac{1}{\mathcal{H}_{t}}\,,
    \label{eq:degradation-function-m}
\end{align}
because $g'(0)=-m$.
The above equation shows that $g(d)$ will be completely determined once a specific expression for $\mathcal{H}_{t}$ is derived.

\begin{figure}[htbp]
    \centering
    \includegraphics[width=0.45\textwidth]{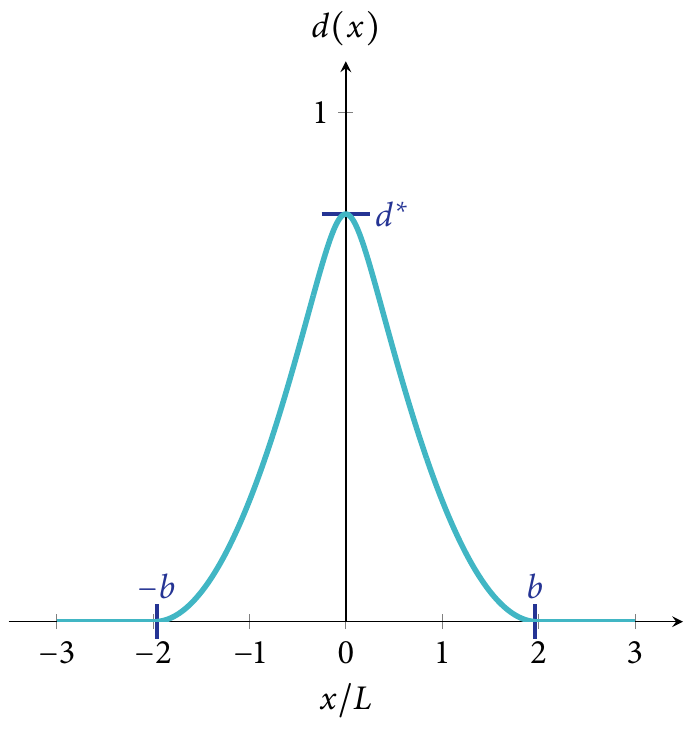}
    \caption{\revised{Definitions of $d^{*}$ and $b$ for a phase-field distribution along the interface normal direction.}}
    \label{fig:d-plot}
\end{figure}

Next, to derive $\mathcal{H}_{t}$, consider a monotonic slip process from the peak shear strength.
Noting that $\tau_{\bulk}=G\gamma$ and $\tau_{p}=G\gamma_{p}$, we can expand $\mathcal{H}_{\slip}$ defined in Eq.~\eqref{eq:crack-driving-force} as follows:
\begin{align}
    \mathcal{H}_{\slip}
    &= \int^{\gamma}_{\gamma_p} (\tau_\bulk - \tau_r)\, \od \gamma \nonumber \\
    &= \int^{\gamma}_{\gamma_p} (G\gamma - \tau_r)\, \od \gamma \nonumber \\
    &= \dfrac{G}{2}[\gamma^2 - \gamma_p^2] - \tau_r [\gamma - \gamma_p] \nonumber \\
    &= \dfrac{G}{2}[\gamma^2 - (\tau_p/G)^2] - \tau_r [\gamma - (\tau_p/G)] \nonumber \\
    &= \dfrac{1}{2G}\left(\tau^2_\bulk - \tau^2_p \right) - \dfrac{\tau_r}{G}\left(\tau_\bulk - \tau_p \right) \nonumber \\
    &= \dfrac{1}{2G}\left(\tau_\bulk + \tau_p - 2\tau_r \right)\left(\tau_\bulk - \tau_p \right).
    \label{eq:H-slip-1d}
\end{align}
Also, rearranging Eq.~\eqref{eq:tau-1d} gives
\begin{align}
    \tau_\bulk = \dfrac{\tau - \tau_r}{g(d)} + \tau_r\,.
    \label{eq:tau-bulk-1d}
\end{align}
Then we insert Eq.~\eqref{eq:tau-bulk-1d} into Eq.~\eqref{eq:H-slip-1d} and get
\begin{align}
    \mathcal{H}_{\slip}
    &= \dfrac{1}{2G}\left(\dfrac{\tau - \tau_r}{g(d)} + \tau_p - \tau_r \right)\left(\dfrac{\tau - \tau_r}{g(d)} + \tau_r - \tau_p \right) \nonumber \\
    &= \dfrac{1}{2G}\left[\left(\dfrac{\tau - \tau_r}{g(d)} \right)^2 - \left(\tau_p - \tau_r\right)^2 \right]\,.
\end{align}
Substituting the above equation into Eq.~\eqref{eq:crack-driving-force}, we obtain an alternative expression for $\mathcal{H}$ as
\begin{align}
    \mathcal{H} = \Lambda + \dfrac{1}{2G}\left(\dfrac{\tau - \tau_r}{g(d)} \right)^2 \,, \quad \text{where} \;\;
    \Lambda := \mathcal{H}_{t} - \dfrac{1}{2G}\left(\tau_p - \tau_r\right)^2\,.
    \label{eq:crack-driving-force-v2}
\end{align}
We now consider the balance of microforce in the slip direction, which is essentially a 1D simple shear problem.
For this 1D setting, the microforce balance equation can be written as
\begin{align}
    \dfrac{3\sfe}{8L}\left(2L^2\dfrac{\od ^2 d}{\od x^2} -1\right) - g'(d)\left[\Lambda + \dfrac{1}{2G}\left(\dfrac{\tau - \tau_r}{g(d)} \right)^2\right] = 0\,,
\end{align}
where $\mathcal{H}$ is substituted by Eq.~\eqref{eq:crack-driving-force-v2}.
To solve this equation for $\tau$, we adopt the solution procedure of Geelen~\etal~\cite{Geelen2019} for a 1D phase-field equation for fracture.
Multiplying $\od d / \od x$ to the above equation gives
\begin{align}
    \dfrac{\od }{\od x} \left\{\dfrac{3\sfe}{8L}\left(L^2 \left(\dfrac{\od d}{\od x} \right)^2 -d \right) - g(d)\left[\Lambda - \dfrac{1}{2G}\left(\dfrac{\tau - \tau_r}{g(d)} \right)^2\right]\right\} = 0\,.
    \label{eq:micro-force-1d}
\end{align}
To integrate this equation, consider the following boundary conditions:
\begin{align}
    d(d^{*},b) = 0\,, \quad
    \dfrac{\od d}{\od x}(d^{*},b) = 0\,.
\end{align}
Here, $b$ denotes the half width of the interface region along the interface normal direction, which may be related to the maximum value of the phase-field variable $d^{*}$ in the interface region.
\revised{See Fig.~\ref{fig:d-plot} for illustration.}
Integrating Eq.~\eqref{eq:micro-force-1d} with the above boundary conditions gives
\begin{align}
    \dfrac{3\sfe}{8L}\left[L^2\left(\dfrac{\od d}{\od x} \right)^2  - d\right] + [1 - g(d)]\Lambda  + \dfrac{1-g(d)}{2Gg(d)}(\tau - \tau_r)^2 = 0 \,.
    \label{eq:micro-force-1d-v2}
\end{align}
To further simplify the equation, recall that the distribution of the phase field must be symmetric across $[-b,b]$.
This symmetry consideration gives the following conditions:
\begin{align}
    d(d^{*},0) = d^{*} \,, \quad
    \dfrac{\od d}{\od x} (d^{*},0) = 0\,.
\end{align}
Using these additional conditions, we rewrite Eq.~\eqref{eq:micro-force-1d-v2} as
\begin{align}
    - \dfrac{3\sfe}{8L}d^{*} + [1 - g(d^{*})]\Lambda + \dfrac{1-g(d^{*})}{2Gg(d^{*})}(\tau - \tau_r)^2 = 0\,,
\end{align}
which is a function of $d^{*}$.
Then, we can express $\tau$ as
\begin{align}
    \tau &= \tau_r + \sqrt{\left\{\dfrac{3\sfe}{8L}d^{*} - [1 - g(d^{*})]\Lambda \right\}\dfrac{2G g(d^{*})}{1-g(d^{*})}}\,.
    \label{eq:tau-derived-v1}
\end{align}
According to Eqs.~\eqref{eq:degradation-function-form} and~\eqref{eq:degradation-function-m}, one can see that
\begin{align}
    \dfrac{g(d^{*})}{1 - g(d^{*})}
    = \dfrac{(1 - d^{*})^n}{m d^{*} (1 + pd^{*})} \,
    = \dfrac{8L\mathcal{H}_{t}}{3\sfe}\dfrac{(1 - d^{*})^n}{d^{*} (1 + pd^{*})} \, .
\end{align}
Inserting the above equation into Eq.~\eqref{eq:tau-derived-v1}, we get
\begin{align}
    \tau = \tau_r + \sqrt{\left\{\dfrac{3\sfe}{8L}d^{*} - [1 - g(d^{*})]\Lambda \right\}\dfrac{8L\mathcal{H}_{t}}{3\sfe}\dfrac{2G (1 - d^{*})^n}{d^{*} (1 + pd^{*})}} \,.
    \label{eq:tau-derived-v2}
\end{align}
From Eq.~\eqref{eq:tau-derived-v2}, one can find that any $\Lambda\neq 0$ will render $\tau$ as a function of $L$, which means that the shear stress will be sensitive to the phase-field length parameter.
Therefore, $\Lambda$ should be zero to ensure that the stress--strain behavior will be independent of $L$.
The requirement of $\Lambda=0$ in Eq.~\eqref{eq:crack-driving-force-v2} leads to the conclusion that $\mathcal{H}_{t}$ should take the following form:
\begin{align}
    \mathcal{H}_{t} = \dfrac{1}{2G}(\tau_p - \tau_r)^2\,.
    \label{eq:critical-fracture-energy-final}
\end{align}
It is noted that $\mathcal{H}_{t}$ can be expressed using three prescribed material parameters, namely $G$, $\tau_p$, and $\tau_r$.
To prove the length-insensitivity of the resulting phase-field formulation, we insert Eq.~\eqref{eq:critical-fracture-energy-final} into Eq.~\eqref{eq:tau-derived-v2}.
Then the shear stress becomes expressed as
\begin{align}
    \tau = \tau_r + \sqrt{\dfrac{d^{*}(1-d^{*})^n}{d^{*}(1+pd^{*})}(\tau_p - \tau_r)^2}\,, \label{eq:tau-derived-v3}
\end{align}
where $L$ no longer appears.
One can also see that $p$ in the degradation function~\eqref{eq:degradation-function-form} controls how shear stress evolves during the strain-softening phase.

The above 1D problem can be further used to analyze the phase-field model derived in this work, much like how Geelen \etal~\cite{Geelen2019} analyzed their own phase-field model of cohesive tensile fracture.
In Appendix, we present such analysis of the proposed model in terms of fracture energy dissipation and slip displacement.
\smallskip

\begin{remark}
    The derived expression for the threshold fracture energy, Eq.~\eqref{eq:critical-fracture-energy-final}, is fully consistent with the crack driving force at the tip of a slip surface derived by Palmer and Rice~\cite{Palmer1973}.
    The crack driving force at the tip, Eq. (17) in in Palmer and Rice~\cite{Palmer1973}, can be written in our notation as
    \begin{align}
        \frac{h}{2G}(\tau_p - \tau_r)^2\,,
    \end{align}
    where $h$ is the height of the domain containing a horizontal slip surface under simple shear.
    Because the threshold fracture energy is defined as the crack driving force at peak stress per unit volume, $h=1$ for $\mathcal{H}_{t}$---and this leads to the exactly same equation as Eq.~\eqref{eq:critical-fracture-energy-final}.
    This theoretical consistency further justifies our derivation of Eq.~\eqref{eq:critical-fracture-energy-final}.
\end{remark}
\smallskip

\begin{remark}
    For phase-field models of cohesive fracture, it is known that there exists an upper bound on $L$ which is related to the fracture energy and the threshold fracture energy.
    The upper bound on our $L$ may be found by substituting $\sfe$ and $\mathcal{H}_{t}$ in Eq.~\eqref{eq:critical-fracture-energy-final} into Eq. (27) of Geelen \etal~\cite{Geelen2019} (note that our $L$ equals one half of $L$ in their paper).
    This gives
    \begin{align}
        L \leq \kappa\frac{\sfe G}{(\tau_p - \tau_r)^2}\,,
        \label{eq:L-upper-bound}
    \end{align}
    where $\kappa=3/4(p+2)$ and $\kappa=3/4$ for the quasi-quadratic and quasi-linear degradation functions, respectively.
    Notably, this upper bound is closely related to the size of the process zone in the theory of Palmer and Rice~\cite{Palmer1973}.
    To wit, we combine Eqs.~(39) and (44) of Palmer and Rice~\cite{Palmer1973} and write the size of the process zone as
    \begin{align}
        \omega = \frac{9\pi}{16(1 - \nu)}\frac{\sfe G}{(\tau_p - \tau_r)^2} \,.
        \label{eq:process-zone-size}
    \end{align}
    One can see that the upper bound on $L$, Eq.~\eqref{eq:L-upper-bound}, and the process zone size $\omega$, Eq.~\eqref{eq:process-zone-size}, are commonly proportional to $\sfe G/(\tau_p - \tau_r)^2$.
    We note that similar relations have also been found between the size of the process zone in opening fracture and the upper bound on the length parameter of existing phase-field models of cohesive tensile fracture (\eg~\cite{Geelen2019,Wu2017}).
    This similarity affirms the consistency of our model with phase-field models of cohesive fracture as well as the theory of Palmer and Rice~\cite{Palmer1973}.
\end{remark}
\smallskip

\begin{remark}
    The quasi-quadratic and quasi-linear degradation functions, which are two particular cases of Eq.~\eqref{eq:degradation-function-form}, may give rise to significant differences in the overall material response although both functions share the same $\mathcal{H}_{t}$ and thus ensure insensitivity to $L$.
    To illustrate this aspect, in Fig.~\ref{fig:degradation-comparison-all} we plot how $g(d)$ and $g'(d)$ vary with $d$, for the quasi-quadratic (with $p=1$) and quasi-linear functions with the same $m$ of 20.
    Figure~\ref{fig:degradation-comparison-value} shows that the value of the quasi-quadratic degradation function approaches zero more rapidly than the quasi-linear one, showing $g(d)\approx 0$ already when $d\geq 0.6$.
    Figure~\ref{fig:degradation-comparison-deriv} indicates that $g'(d)$, which is multiplied to $\mathcal{H}$ in the governing equation~\eqref{eq:governing-slip}, evolve very similarly from their common initial value of $g'(0)=-m$.
    Nevertheless, there is a subtle difference in their final values, $g'(1)$, which determines the overall crack driving force term, $-g'(d)\mathcal{H}$, in a fully-fractured region.
    To explain this difference, we write the closed form of $g'(d)$ as
    \begin{align}
        g'(d) = -\dfrac{m(1-d)^{n-1}\left[ (n-2)pd^2 + d(n+2p-1) + 1\right]}{\left[(1-d)^n + md(1+pd) \right]^2}\,.
    \end{align}
    One can see that, whereas $g'(1)=0$ for the quasi-quadratic function in which $n=2$, $g'(1)=-1/m=-(8L\mathcal{H}_{t})/3\sfe < 0$ for the quasi-linear function in which $n=1$.
    This means that when the quasi-linear function is used, the overall crack driving force term in the governing equation~\eqref{eq:governing-slip} does not vanish in a fully-fractured region.
    However, it is usually very small as shown in Fig.~\ref{fig:degradation-comparison-deriv}, because $m \gg 1$ typically.
    The effect of the choice of $g(d)$ will be further discussed in Section~\ref{sec:examples} based on numerical results.
    \begin{figure}[htbp]
        \centering
        \subfloat[]{\includegraphics[width=0.48\textwidth]{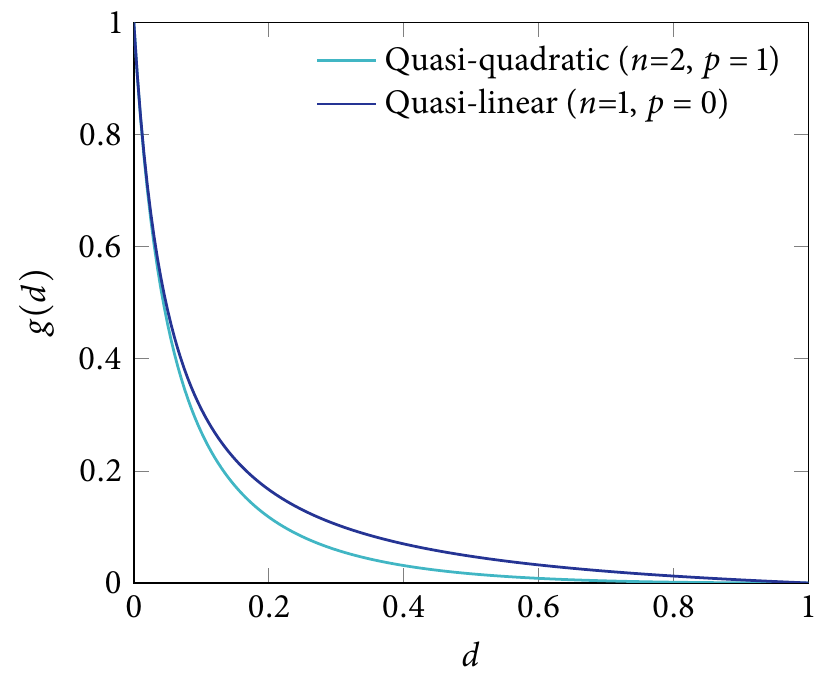} \label{fig:degradation-comparison-value}} \hspace{0.5em}
        \subfloat[]{\includegraphics[width=0.48\textwidth]{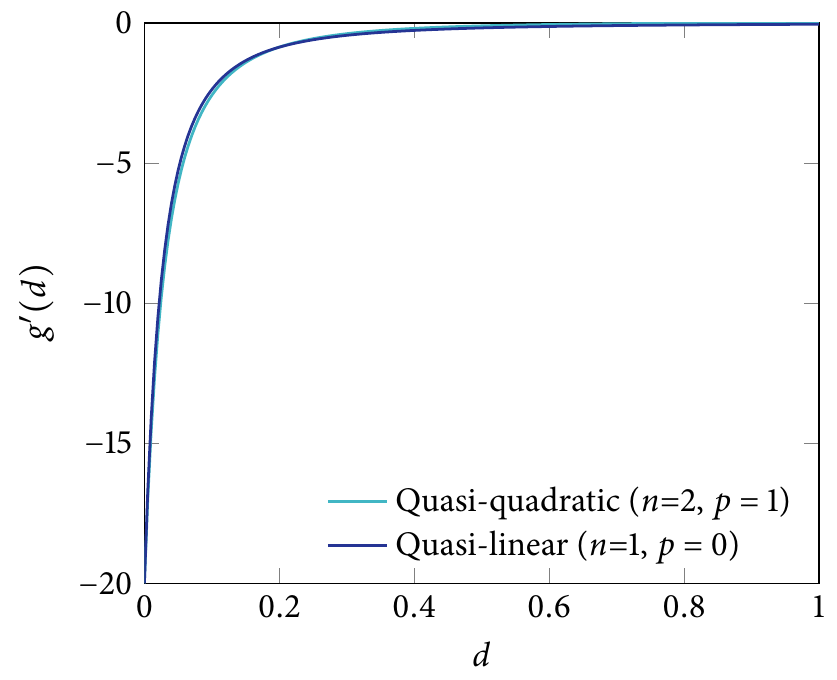} \label{fig:degradation-comparison-deriv}}
        \caption{Variations of (a) the values and (b) the derivatives of the quasi-quadratic and quasi-linear degradation functions with the phase field. See Eq.~\eqref{eq:degradation-function-form} for the definition of $n$ and $p$. For illustration purposes, $m=20$ is assigned in both plots. }
        \label{fig:degradation-comparison-all}
    \end{figure}
\end{remark}
\smallskip

\revised{
\begin{remark}
    As discussed in previous work~\cite{Wu2017,Geelen2019}, the quasi-linear degradation function has a zero ultimate displacement, giving rise to snap-back behavior. For this reason, the quasi-linear function is appropriate only for a particular class of problems in which crack propagates in a stable manner.
    This issue is absent in the quasi-quadratic degradation function which has a finite ultimate displacement.
\end{remark}
}

\subsection{Determination of fracturing direction}
The foregoing phase-field formulation has been developed provided that the direction of fracturing is given.
However, although the direction may be easily determined in simple cases, it is not known \textit{a priori} in general.
In what follows, we devise a method that can be used to determine the fracturing direction in the proposed phase-field formulation.

The premise of our method is that fracturing takes place such that it maximizes energy dissipation.
This premise has been used in eigenfracture/erosion models~\cite{Schmidt2009,Pandolfi2012,Pandolfi2013} to determine a crack path,
and in the mixed-mode phase-field model of Bryant and Sun~\cite{Bryant2018} for the same purpose.
Notably, in the latter work, the fracturing direction in each load step is found as the direction that gives the largest crack driving force.
Adopting this argument, we will also search for the direction that maximizes the crack driving force for frictional shear fracture at hand.

Recall that the crack driving force in our model is given by Eq.~\eqref{eq:crack-driving-force}, where $\mathcal{H}_{t}$ has been derived as Eq.~\eqref{eq:critical-fracture-energy-final}.
Because $\tau_{p}=\tau_{\bulk}$ at the moment of initial fracture, one can see that the crack driving force is proportional to $\tau_{\bulk} - \tau_{r}$ throughout the fracturing process.
We will therefore search for the plane where $\tau_{\bulk} - \tau_r$ attains its maximum value.
For this purpose, let us denote by $\theta$ the inclination angle of a plane to the direction of the major principal stress in compression.
Our task is then to find $\theta$ that maximizes $\tau_{\bulk} - \tau_r$, \ie
\begin{align}
    \theta = \arg\max_{\theta} [\tau_\bulk(\theta) - \tau_r(\theta)]\,.
\end{align}
Adopting the Coulomb criterion, we express $\tau_{r}$ as
\begin{align}
    \tau_r = p_{\cn}\tan\phi_{r}\,,
    \label{eq:residual-strength}
\end{align}
where $p_{\cn}$ is the contact normal pressure acting on a surface, and $\phi_{r}$ is the residual friction angle.
Note that the residual shear strength is considered cohesionless, as suggested by experimental evidence~\cite{Skempton1964}.
Without loss of generality, $p_{\cn}>0$ is assumed in the following discussion.

We then seek to write $\tau_{\bulk}$ and $p_{\cn}$ in terms of $\theta$.\
Let us denote by $s_1$, $s_2$, and $s_3$ the major, intermediate, and minor principal stresses in compression, respectively.
We then focus on the plane constructed by the directions of $s_1$ and $s_3$ in which slip takes places.
In this plane, $\tau_{\bulk}$ and $p_{\cn}$ can be written as
\begin{align}
    \tau_{\bulk} = \dfrac{s_{1} - s_{3}}{2}\sin{2\theta}\,, \quad
    p_{\cn} = \dfrac{s_{1} + s_{3}}{2} - \dfrac{s_{1} - s_{3}}{2}\cos{2\theta}\,.
\end{align}
Inserting the above expressions into Eq.~\eqref{eq:residual-strength} and $\tau_{\bulk} - \tau_{r}$, we get
\begin{align}
    \tau_\bulk - \tau_r
    = -\dfrac{s_{1} + s_{3}}{2}\tan\phi_{r} + \dfrac{s_{1} - s_{3}}{2}\left(  \sin{2\theta} + \cos{2\theta} \tan \phi_{r}\right).
\end{align}
Recall that $\tau_{\bulk} \geq \tau_{r}$ during fracturing as well as that $0\leq\phi_{r}\leq90^{\circ}$ for physically meaningful materials, one can see that $\tau_\bulk - \tau_r$ becomes greater with $h(\theta) := \sin{2\theta} + \cos{2\theta} \tan \phi_{r}$.
Therefore, to find $\theta$ maximizing $h(\theta)$, we solve $h'(\theta) = 2\cos{2\theta} - 2\tan\phi_{r} \sin{2\theta} = 0$ and obtain
\begin{align}
    \theta &= 45^{\circ} - \dfrac{\phi_{r}}{2} \,.
\end{align}
Remarkably, this failure plane angle is the same as that obtained from Mohr--Coulomb theory using the residual strength criterion.
At this point, we also adopt the Mohr--Coulomb criterion for the peak strength, which can be written as
\begin{align}
    \tau_{p} = c + p_{\cn}\tan\phi\,.
    \label{eq:peak-strength}
\end{align}
Here, $c$ is the cohesion strength of the material, and $\phi$ is the peak friction angle.
The peak and residual friction angles are usually distinguished although they are found to be similar in some materials; see~\cite{Skempton1964,Mitchell2005,Paterson2005,Chang2006} for experimental data on soils and rocks.

Once $\theta$ is determined, $\tensor{n}$ and $\tensor{m}$ of the (potential) failure plane may be obtained according to Anderson's fault theory~\cite{Anderson1951}.
First, because $\tensor{n}$ is orthogonal to the direction of $s_2$, it can be calculated by rotating the unit vector in the direction of $\stress_{1}$ along the direction of $s_{2}$ by $90^{\circ} - \theta$ clockwise.
Using Rodrigues' formula for vector rotation, $\tensor{n}$ is computed as
\begin{align}
    \tensor{n} = \tensor{a}_{1}\sin{\theta} + \left(\tensor{a}_{2} \times \tensor{a}_{1} \right) \cos \theta\,,
\end{align}
where $\tensor{a}_{1}$ and $\tensor{a}_{2}$ denote the unit vectors in the directions of $s_1$ and $s_2$, respectively.
Next, because $\tensor{m}$ is mutually orthogonal to $\tensor{n}$ and $\tensor{a}_{2}$, it is determined as
\begin{align}
    \tensor{m} = \tensor{a}_\mathrm{2} \times \tensor{n}\,.
\end{align}
Then we can also calculate $\tensor{\alpha}=(\tensor{n}\dyad\tensor{m} + \tensor{m}\dyad\tensor{n})$.
\smallskip

\begin{remark}
    If desired, alternative expressions for the peak and/or residual shear strengths can be employed within the proposed phase-field modeling framework.
    For example, a nonlinear strength criterion (\eg~Atkinson~\cite{Atkinson2007} for overconsolidated clays or Shen \etal~\cite{Shen2018} for rocks) may be used for the peak strength in lieu of Eq.~\eqref{eq:peak-strength}.
    In this work, the Mohr--Coulomb strength criterion are selected to minimize the number of material parameters and to compare our numerical results with others in the literature produced by other numerical methods using the same strength criterion.
\end{remark}

\section{Discretization and algorithms}
\label{sec:discretization}

This section presents discrete formulations and algorithms to apply the proposed phase-field model to simulate shear fracture processes in geologic materials.
To fully focus on the topic at hand---frictional shear cracks---we will exclude the case of open cracks.
Note that the open crack case can be augmented straightforwardly by checking $\strain_{\cn} > 0$ as discussed in the beginning of the previous section.

\subsection{Problem statement}
We begin by stating an initial--boundary-value problem based on the proposed phase-field formulation.
Before delving into the problem statement, we note that the governing equations for intact and stick conditions, Eqs.~\eqref{eq:governing-intact} and~\eqref{eq:governing-stick} respectively, can be unified because $\tstress=\tstress_{\bulk}$ and fracturing does not occur under both conditions.
Therefore, in what follows, we consider two modes, namely the non-slip mode (intact and stick) and the slip mode, and distinguish them using the yield function, $f := |\tau| - \tau_{\yield}$, by setting $\tau_{\yield}=\tau_{p}$ for an intact material and $\tau_{\yield}=\tau_{r}$ for a damaged material.

The strong form of the problem can then be stated as follows. Given $\hat{\tensor{u}}$, $\hat{\tensor{t}}$, $\tensor{u}_{0}$, and $d_{0}$, find $\tensor{u}$ and $d$ such that
\begin{align}
  \diver\,\tstress + \rho\tensor{g} &= \tensor{0} \quad\text{in}\;\;\Omega\times\mathbb{T}\,, \\
  -g'(d)\mathcal{H}^{+}  + \dfrac{3\sfe}{8L}\left(2L^2\diver \grad{d} - 1\right) &= 0 \quad\text{in}\;\;\Omega\times\mathbb{T}\,,
\end{align}
where
\begin{align}
    \tstress
    = \left\{\begin{array}{ll}
        \tstress_{\bulk} & \text{if}\;\; f<0\,,  \\ [0.5em]
        \tstress_{\bulk} - [1 - g(d)]\tau_{\bulk}\tensor{\alpha} + [1 - g(d)]\tau_{r}\tensor{\alpha} & \text{if} \;\; f = 0 \,,
    \end{array}\right.
    \label{eq:stress-combined}
\end{align}
and
\begin{align}
    \mathcal{H}^{+}
    = \left\{\begin{array}{ll}
        \displaystyle\max_{t\in[0,t]} \mathcal{H}^{+}(t) & \text{if}\;\; f<0\,, \\ [0.5em]
        \mathcal{H}_{t} + \mathcal{H}_{\slip} & \text{if} \;\;  f=0\,,
    \end{array}\right.
    \label{eq:crack-driving-force-combined}
\end{align}
subjected to the boundary conditions
\begin{align}
  \tensor{u} = \hat{\tensor{u}} \quad&\text{on}\;\;\partial_{u}\Omega\times\mathbb{T}\,,\\
  \tensor{\sigma}\cdot\tensor{\upsilon} = \hat{\tensor{t}} \quad&\text{on}\;\;\partial_{t}\Omega\times\mathbb{T}\,, \\
  \grad{d}\cdot\tensor{\upsilon} = 0 \quad&\text{on}\;\;\partial\Omega\times\mathbb{T}\,,
\end{align}
and the initial conditions
\begin{align}
    \tensor{u}|_{t=0} &= \tensor{u}_{0} \quad\text{in}\;\;\overline{\Omega}\times\mathbb{T}\,, \\
    d|_{t=0} &= d_{0} \quad\text{in}\;\;\overline{\Omega}\times\mathbb{T}\,.
\end{align}
Here, $\hat{\tensor{u}}$ and $\hat{\tensor{t}}$ are prescribed boundary conditions of displacement and traction vectors, respectively, $\tensor{\upsilon}$ is the outward unit normal vector at the domain boundary, and $\overline{\Omega}:=\overline{\Omega\cup\partial\Omega}$.
It is also noted that $\mathcal{H}^{+}$ for intact and stick conditions can be combined as Eq.~\eqref{eq:crack-driving-force-combined} because the initial $\mathcal{H}^{+}$ for an intact material is $\mathcal{H}_{t}$.

\subsection{Finite element discretization}
We use the standard Galerkin finite element method to discretize the two-field governing equations at hand.
To begin, we introduce trial solution spaces for $\tensor{u}$ and $d$ as
\begin{align}
  \mathcal{S}_{u} &:= \{\tensor{u} \;\vert\; \tensor{u} \in H^{1}, \; \tensor{u}=\hat{\tensor{u}} \;\;\text{on} \;\; {\pd_{u}\Omega} \}, \\
  \mathcal{S}_{d} &:= \{d \;\vert\; d \in H^{1}\},
\end{align}
where $H^{1}$ denotes a Sobolev space of order one.
Accordingly, we define weighting function spaces for the two fields as
\begin{align}
  \mathcal{V}_{u} &:= \{\tensor{\eta} \;\vert\; \tensor{\eta} \in H^{1}, \; \tensor{\eta}=\tensor{0} \;\;\text{on} \;\; {\pd_{u}\Omega} \}, \\
  \mathcal{V}_{d} &:= \{\varphi \;\vert\; \varphi \in H^{1} \}\,.
\end{align}
Through the weighted residual procedure, we obtain the following two variational equations:
\begin{align}
  - \int_{\Omega} \symgrad\tensor{\eta}:\tensor{\sigma}\,\od V
  + \int_{\Omega} \rho \tensor{\eta}\cdot\tensor{g}\,\od V
  + \int_{\pd_{t}{\Omega}} \tensor{\eta}\cdot\hat{\tensor{t}}\,\od A &= 0\,, \label{eq:var-mom} \\
  \int_{\Omega} \varphi g'(d)\mathcal{H}^{+}\, \od V
  + \int_{\Omega} \frac{3\sfe}{8L}\left(2L^{2}\grad{\varphi}\cdot\grad{d} + \varphi\right) \od V &= 0 \,. \label{eq:var-phasefield}
\end{align}
We then discretize these variational equations using mixed $\mathbb{Q}_{1}$/$\mathbb{Q}_{1}$ finite elements.
The procedure for this discretization is fairly standard, so it will be omitted for brevity.

To solve the discrete versions of Eqs.~\eqref{eq:var-mom} and~\eqref{eq:var-phasefield}, we use the sequential update scheme proposed by Miehe \etal~\cite{Miehe2010b}, which has been dominant in phase-field modeling of fracture.
Specifically, in the slip mode ($f=0$), we solve Eq.~\eqref{eq:var-mom} calculating $\tensor{\sigma}$ with a fixed $d$, and Eq.~\eqref{eq:var-phasefield} evaluating $\mathcal{H}_{\slip}$ with a fixed $\tensor{u}$.
Usually, this sequential scheme does not much compromise the solution accuracy because the load increment should be very small during the process of slip.
It is also noted that Eq.~\eqref{eq:var-mom} and~\eqref{eq:var-phasefield} are decoupled in the non-slip mode ($f<0$).

\subsection{Material update algorithm}
Unlike standard phase-field models of fracture, the proposed model requires a non-trivial algorithm for updating variables at a material/quadrature point (\eg~stress and crack driving force) because multiple responses are possible depending on the contact condition (or lack thereof).
Given that the yield function $f$ is used to distinguish between slip and non-slip modes,
we design a material update algorithm adopting a predictor--corrector format from a standard return mapping procedure in plasticity.

The procedure designed to update material-level variables from time $t_{n}$ to $t_{n+1}$ is summarized in Algorithm~\ref{algo:material-update}.
For notational simplicity, variables at $t_{n+1}$ are denoted without an additional subscript, while those at $t_{n}$ is denoted as $(\cdot)_{n}$.
Notably, for the stress update under a slip condition, we evaluate $\tau_{r}$ with the contact normal pressure at the previous time step, $(p_{\cn})_{n}$.
This semi-implicit update greatly simplifies the expression for the stiffness tensor, without significant compromise to the solution accuracy because the step size is maintained small during a slip process.
The increment of the slip-induced crack driving force, $\dot{\mathcal{H}}_{\slip} = (\tau_{\bulk}-\tau_{r})\dot{\gamma}$, is evaluated implicitly with $\tau_{\bulk}$ and $\tau_{r}$ at $t_{n+1}$.
\begin{algorithm}[htbp]
    \setstretch{1.1}
    \caption{Material point update procedure for phase-field modeling of frictional shear fracture}
    \begin{algorithmic}[1]
        \Require $\tstrain$, $d$, $\tensor{n}$, and $\tensor{\alpha}$ at $t_{n+1}$.
        \Ensure $\tstress$, $\mathbb{C}$, $\mathcal{H}^{+}$, and $g(d)$ at $t_{n+1}$.

        \State Calculate  $\tstress_{\bulk} = \mathbb{C}_{\bulk}:\tstrain$, $\tau_{\bulk} = (1/2)\tstress_{\bulk}:\tensor{\alpha}$, and $p_{\cn} := -\tstress_{\bulk}:(\tensor{n}\dyad\tensor{n})$.

        \State Set $\tau_{\yield} = c + p_{\cn}\tan\phi$ if $d=0$; $\tau_{\yield} = p_{\cn}\tan\phi_{r}$ otherwise.

        \State Evaluate $f = |\tau_{\bulk}| - \tau_{\yield}$.

        \If {$f < 0$}
            \State Update $\tstress = \tstress_{\bulk}$ and $\mathbb{C} = \mathbb{C}_{\bulk}$.
            \If {$d = 0$}
                \State Intact condition.
                \State Update $\mathcal{H}^{+} = \mathcal{H}_{t}$ by substituting $\tau_p = c + p_{\cn} \tan\phi$ and $\tau_r = p_{\cn} \tan\phi_{r}$ into Eq.~\eqref{eq:critical-fracture-energy-final}.
                \State Update $g(d)$ by calculating $m$ as Eq.~\eqref{eq:degradation-function-m} with the updated $\mathcal{H}_{t}$.
            \Else
                \State Stick condition.
                \State Set $\mathcal{H}^{+} = \displaystyle\max_{t\in[0,t_{n}]} \mathcal{H}^{+}(t)$.
            \EndIf
        \Else
            \State Slip condition.
            \State Update $\tstress = \tstress_{\bulk} - [1 - g(d)][\tau_{\bulk} - (\tau_{r})_{n}]\tensor{\alpha}$, where $\tau_{r,n} := (p_{\cn})_{n} \tan\phi_{r}$.
            \State Update $\mathbb{C} = \mathbb{C}_{\bulk} - [1 - g(d)]G(\tensor{\alpha}\dyad\tensor{\alpha})$
            \State Update $\mathcal{H}^{+} = (\mathcal{H}^{+})_{n} + (\tau_{\bulk} - \tau_r)\Delta \gamma$, where $\tau_r = p_{\cn} \tan\phi$ and $\Delta\gamma := (\tstrain - \tstrain_{n}):\tensor{\alpha}$.
        \EndIf
  \end{algorithmic}
  \label{algo:material-update}
\end{algorithm}


\section{Numerical examples}
\label{sec:examples}

Three numerical examples are presented in this section to verify and investigate the proposed phase-field model.
\revised{
The first two examples are designed to verify the model, with an emphasis on the sensitivity to the phase-field length parameter, $L$.
In the third example, the phase-field model is applied to simulate the initiation and propagation of a slip surface in a field-scale slope failure problem.
By default, we use the quasi-quadratic degradation function, setting $n=2$ and $p=1$ in Eq.~\eqref{eq:degradation-function-form}.
The reason is that the quasi-linear degradation function gives rise to snap-back behavior as well as a non-zero crack driving force in a fully-damaged region, as discussed previously at the end of Section~\ref{sec:formulation}.
So only the quasi-quadratic degradation function is appropriate for the first two examples which focus on verification of the proposed model.
However, in the last example, both the quasi-quadratic and quasi-linear degradation functions are used to demonstrate how they lead to significantly different responses of the same system.
}

The results of the numerical examples have been produced by our in-house finite element code used in previous work (\eg~\cite{Choo2015,Choo2016,Borja2016,Zhang2019}).
The code relies heavily on open source scientific computing libraries including the \texttt{deal.II} finite element library~\cite{Bangerth2007,dealII90}, \texttt{p4est} mesh handling library~\cite{Burstedde2011}, and the \texttt{Trilinos} project~\cite{Heroux2012}.
In all numerical examples, plane strain conditions are assumed and quadrilateral elements are used.

\subsection{Long shear apparatus}
\begin{figure}[htbp]
    \centering
    \includegraphics[width=0.8\textwidth]{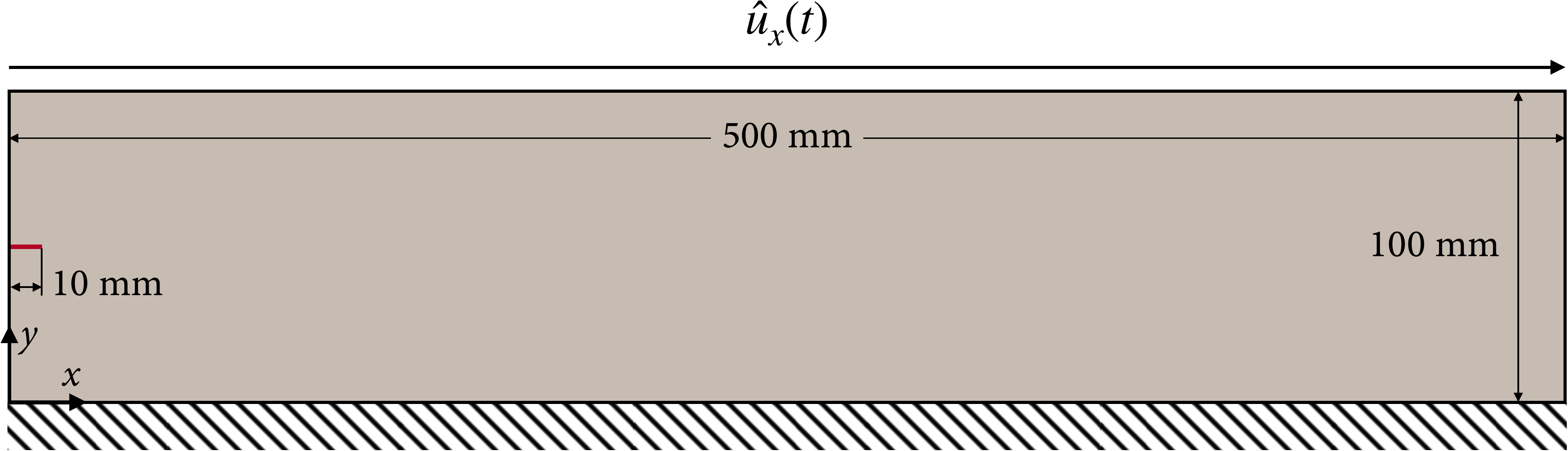}
    \caption{Long shear apparatus: problem geometry and boundary conditions. The horizontal line at the left boundary denotes the initial slip surface in the domain.}
    \label{fig:long-shear-setup}
\end{figure}

The purpose of our first example is to verify the proposed model numerically with respect to its consistency with Palmer and Rice's theory~\cite{Palmer1973}.
For this purpose, we simulate growth of a slip surface in a long shear apparatus, for which Palmer and Rice derived their formulation~\cite{Palmer1973}.
Figure~\ref{fig:long-shear-setup} depicts the setup of the problem, which is the same as that illustrated in Fig. 3 of Palmer and Rice~\cite{Palmer1973} except that specific dimensions are assigned to the domain herein.
The domain is 500-mm long and 100-mm tall, and it has a 10-mm notch in the middle of the left boundary which serves as an initial slip surface.
As in the original problem, the top boundary is displaced horizontally, whereas the bottom boundary is fixed.
The two lateral boundaries are fixed vertically.
Gravity is ignored.
\revised{Because Palmer and Rice's theory~\cite{Palmer1973} considers 1D slip surface propagation, we restrict the propagation direction to be horizontal.}

Considering that the original paper of Palmer and Rice~\cite{Palmer1973} is concerned with overconsolidated clays,
we assign material parameters similar to those of overconsolidated clays in the literature (see~\cite{Skempton1964,Manica2018} for example).
They are: the cohesion strength $c=40$ KPa, the peak and residual friction angles $\phi=\phi_{r}=15^{\circ}$, the shear modulus $G=10$ MPa, and Poisson's ratio $\nu = 0.3$.
We set $\sfe=30$ J/m$^2$ by estimating the shear fracture energy of a laboratory-scale stiff clay studied in M\'{a}nica \etal~\cite{Manica2018} using the procedure described in Wong~\cite{Wong1982a}.

To examine the sensitivity of the model to the phase-field length parameter, we consider three cases of $L$, namely, 2 mm, 4 mm, and 8 mm.
By default, the fracture path is discretized with uniform elements whose size $h$ satisfies $L/h=20$.
Using the case of $L=2$ mm, we further check the mesh sensitivity of the model by two coarser discretization levels, namely, $L/h=5$ and 10.
The initial slip surface is generated by assigning large crack driving forces at the quadrature points nearby, as done by Borden \etal~\cite{Borden2012}.
We also initialize vertical compressive normal stress of $149$ kPa at all quadrature points, which gives $\tau_{p}=80$ kPa and $\tau_{r}=40$ kPa.
The simulation is conducted applying a constant displacement rate of 0.01 mm per load step to the top boundary.

\begin{figure}[htbp]
    \centering
    \includegraphics[width=0.55\textwidth]{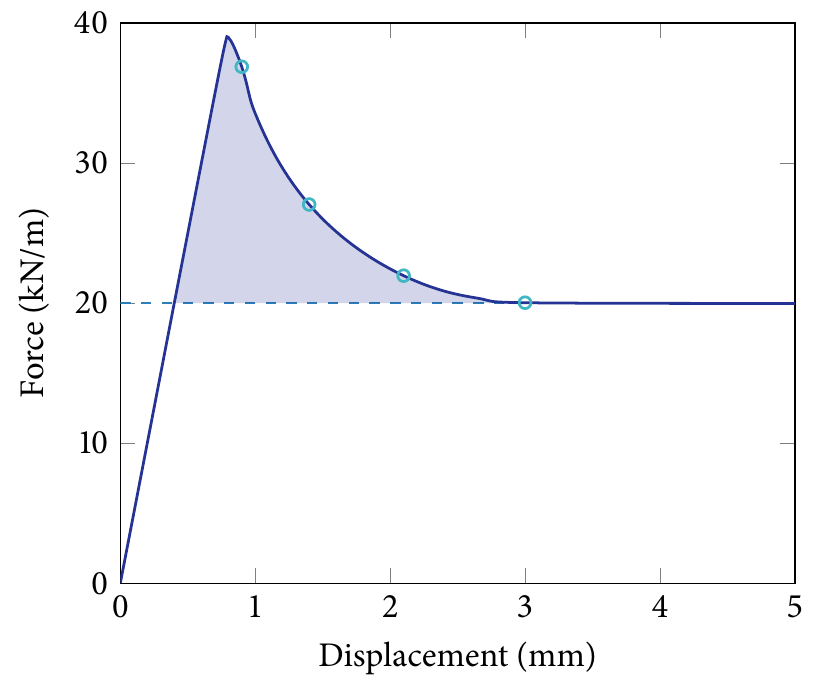}
    \caption{Long shear apparatus: the force--displacement curve with $L=2$ mm and $L/h=20$. The shaded area is the energy dissipated by fracturing (\cf~Fig.~\ref{fig:palmer-rice-illustration}), which is very close to the theoretical value. The open circles denote the points at which the phase-field distributions are drawn in Fig.~\ref{fig:long-shear-contours}.}
    \label{fig:long-shear-verification}
\end{figure}

Figure~\ref{fig:long-shear-verification} presents the force--displacement curve when $L=2$ mm and $L/h=20$.
Firstly, it is found that the model not only well reproduces the characteristic behavior of a quasi-brittle geomaterial---softening from the peak strength to the residual strength---but also provides quantitatively accurate results in that the residual load, 20 kN, equals the residual shear strength times the width of the domain.
The softening curve also resembles those of opening fracture problems simulated by Geelen \etal~\cite{Geelen2019} using the same type of $g(d)$.
Note that the peak load is slightly less than 40 kN because the peak shear strength is not mobilized simultaneously along the potential slip surface.
This is exactly why the use of the peak shear strength characterized from laboratory tests must underestimate the strength of a field-scale geotechnical system, see Skempton~\cite{Skempton1964}.
Also importantly, the fracture energy estimated from the force--displacement curve---the shaded area in the figure---is very close to its theoretical value, showing difference less than 0.03\% from the shear fracture energy times the fractured area ($14.7$ J).
All these observations affirm the consistency of the proposed phase-field model with Palmer and Rice's theory~\cite{Palmer1973}.

\begin{figure}[htbp]
    \centering
    \includegraphics[width=0.8\textwidth]{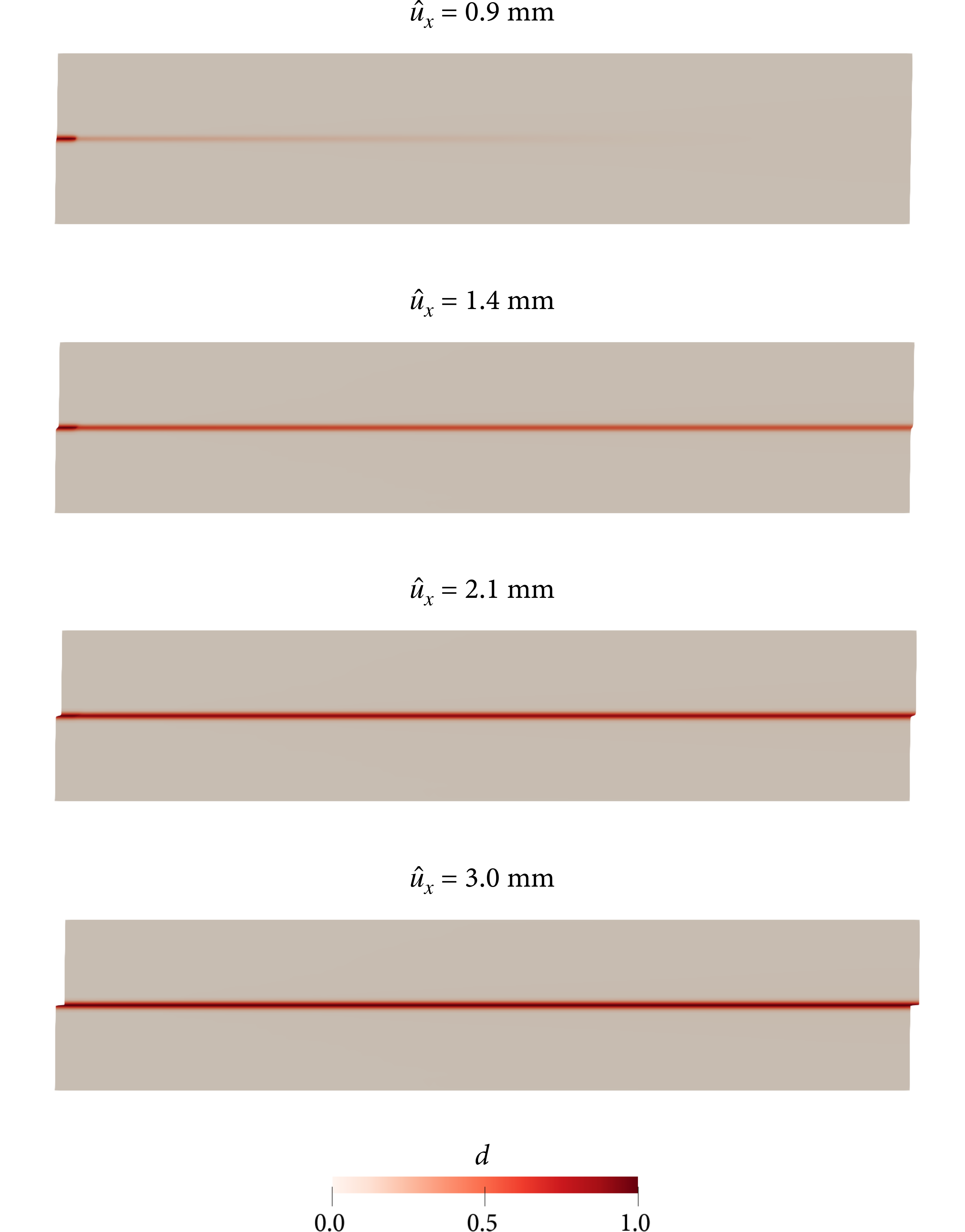}
    \caption{Long shear apparatus: phase-field distributions at the four instances marked in Fig.~\ref{fig:long-shear-verification}. The domain is scaled by the displacement field with a factor of 2.}
    \label{fig:long-shear-contours}
\end{figure}

Figure~\ref{fig:long-shear-contours} plots the distributions of the phase field at the four instances marked in Fig.~\ref{fig:long-shear-verification}.
After the peak load ($\hat{u}_{x}=0.9$ mm), the process zone emerges and propagates much like the conceptual illustration in Fig.~\ref{fig:palmer-rice-illustration}.
It then reaches the right end and intensifies during the softening stage, as can be seen from the phase-field distributions at $\hat{u}_{x}=1.4$ mm and $\hat{u}_{x}=2.1$ mm.
Once the slip surface has fully developed ($\hat{u}_{x}=3.0$ mm), the domain has been completely split into two parts, of which the upper one slides over the lower one.
In this final stage, the shear stress mobilized along the slip surface is constant throughout and equal to the residual shear strength.

\begin{figure}[htbp]
    \centering
    \subfloat[]{\includegraphics[width=0.48\textwidth]{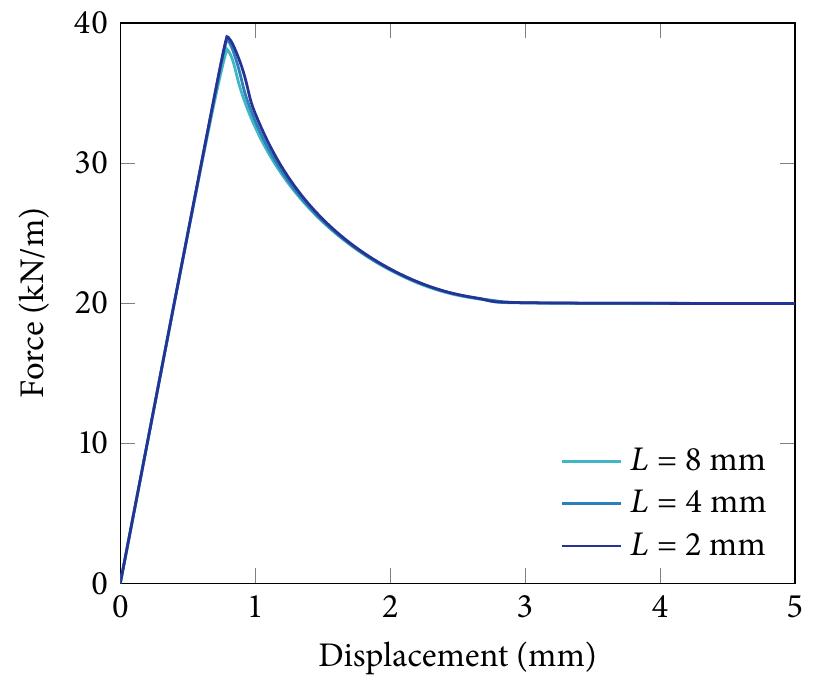} \label{fig:long-shear-L}} \hspace{0.5em}
    \subfloat[]{\includegraphics[width=0.48\textwidth]{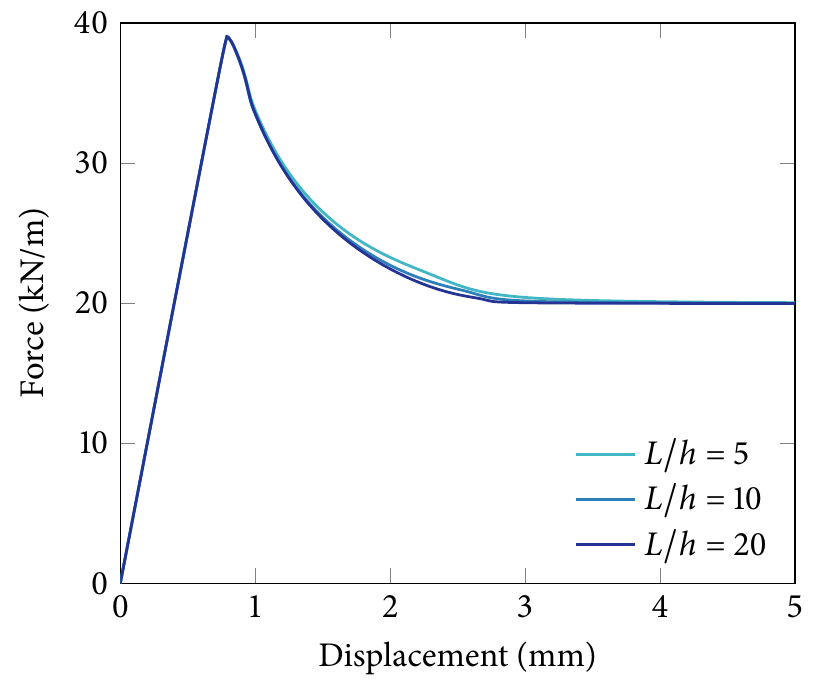} \label{fig:long-shear-h}}
    \caption{Long shear apparatus: force--displacement curves with different phase-field length parameters ($L$) and element sizes ($h$). (a) $L$ is varied fixing $L/h=20$. (b) $h$ is varied fixing $L=2$ mm.}
\end{figure}

Using this fundamental problem, we investigate the sensitivity of the proposed model to the phase-field length parameter ($L$) and the element size ($h$).
The force--displacement curves of the three cases of $L$ with a fixed $L/h$, and the three cases of $L/h$ with a fixed $L$, are presented in Figs.~\ref{fig:long-shear-L} and~\ref{fig:long-shear-h}, respectively.
These curves confirm that the model is virtually insensitive to both $L$ and $h$.
It is noted that the mesh insensitivity is a common feature of phase-field formulations as they are inherently nonlocal, while the length insensitivity is a particular achievement of this work by deriving specific forms of $g(d)$ and $H_{t}$.
Therefore, in the next example, we shall focus on the length insensitivity of the phase-field model.

\subsection{Biaxial compression}
Our second example simulates biaxial compression tests on laboratory-scale specimens under different confining pressures, until the specimens fail by faulting.
The main objective of this example is to probe the ability of the model to capture the pressure dependence of the peak and residual strengths without much sensitivity to the phase-field length parameter.
As explained in Introduction, attaining this ability is a major challenge addressed in this work because the existing phase-field models of cohesive fracture considered pressure-independent, constant peak strengths only.

\begin{figure}[htbp]
    \centering
    \includegraphics[width=0.4\textwidth]{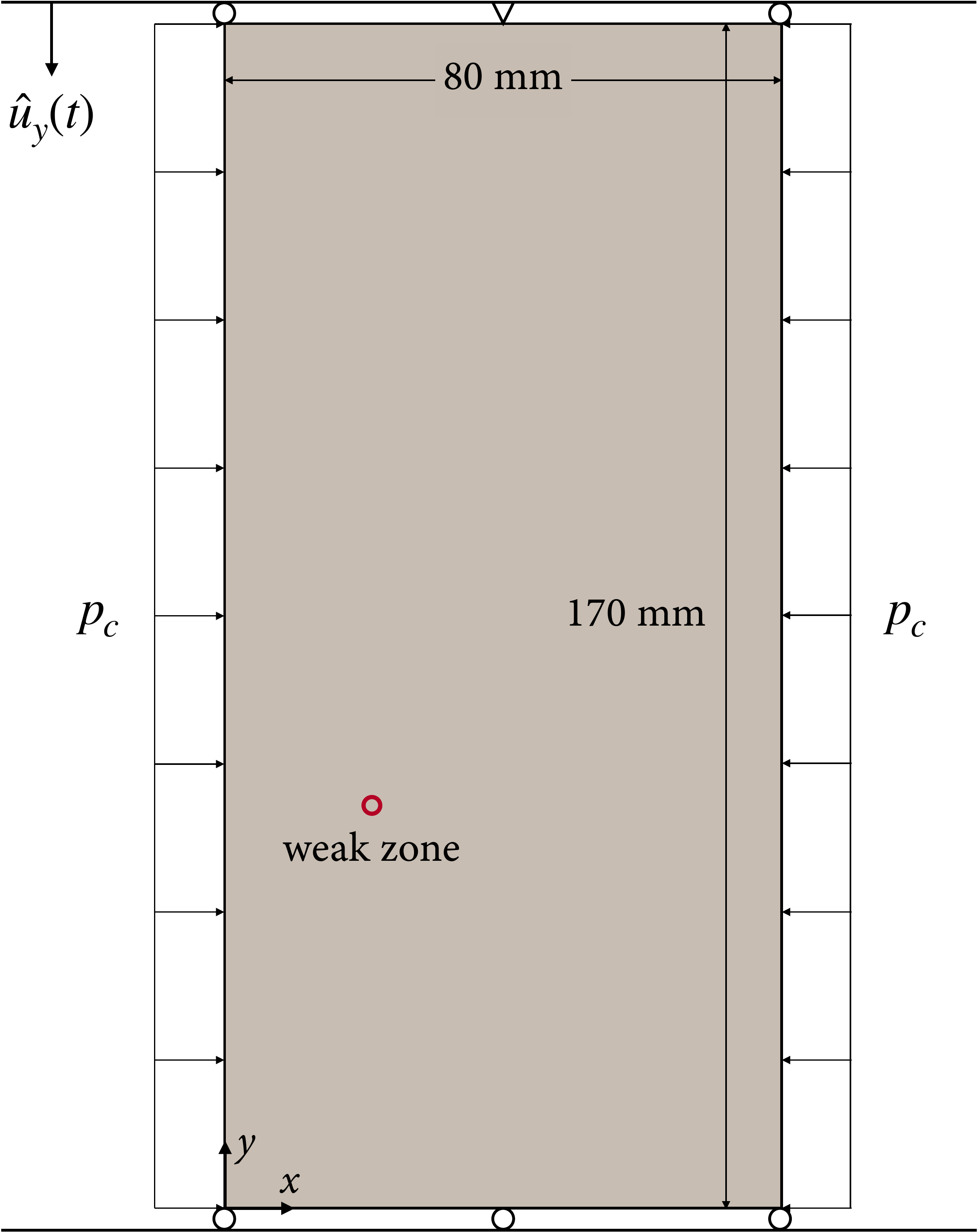}
    \caption{Biaxial compression: problem geometry and boundary conditions. The circle denotes the weak zone introduced as an imperfection triggering a unique shear fracture.}
    \label{fig:biaxial-setup}
\end{figure}

The setup of the problem is illustrated in Fig.~\ref{fig:biaxial-setup}.
The domain is a 80-mm wide and 170-mm tall rectangular specimen, of which the top boundary is vertically compressed by a prescribed displacement.
The top and bottom boundaries are supported by rollers throughout,
 except at the top middle which is attached to a pin for stability.
The two lateral boundaries are subjected to confining pressure ($p_c$) which remains constant during the compression test.
The material properties are the same as those of the previous example,
except in the circular weak zone where $\sfe$ is reduced to 20 J/m$^2$ to trigger a unique shear fracture.
The location of the weak zone is also drawn in Fig.~\ref{fig:biaxial-setup}.
Gravity is neglected again.
To examine whether the proposed model well incorporates the pressure dependence of the peak and residual strengths in a  length-insensitive manner,
we consider three cases of $p_{c}$, namely, 50 kPa, 100 kPa, and 200 kPa, and repeat each case with two values of $L$, namely, 1 mm and 2 mm.
The domain is discretized by a structured mesh in which elements along the potential fracturing region are locally refined to satisfy $L/h=20$.
The compression rate is controlled adaptively, such that it becomes as small as 0.001 mm per load step during the softening process.

\begin{figure}[htbp]
    \centering
    \includegraphics[width=0.55\textwidth]{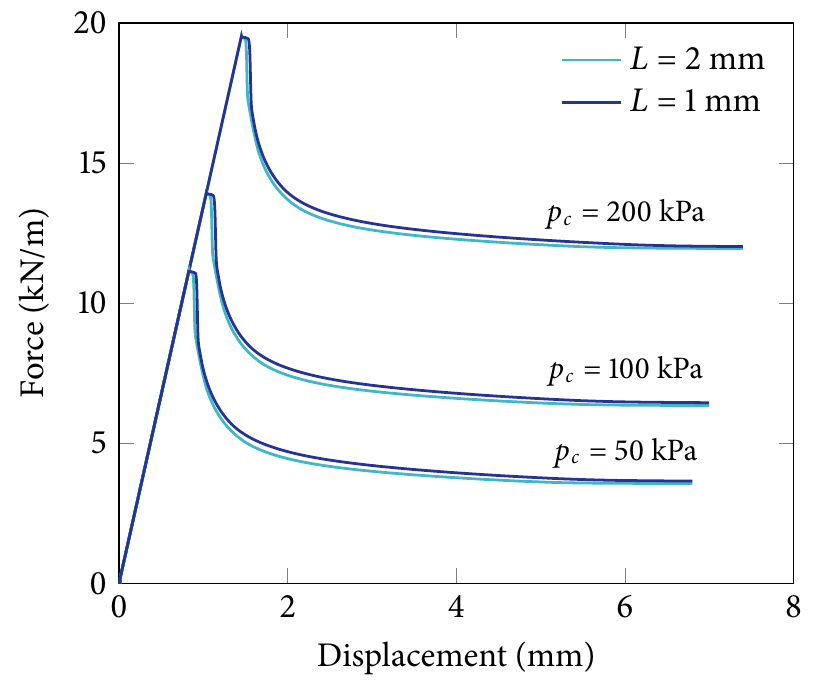}
    \caption{Biaxial compression: force--displacement curves at different confining pressures ($p_c$).
    Each case of confining pressure is simulated with two different phase-field length parameters ($L$).}
    \label{fig:biaxial-L}
\end{figure}

The force--displacement curves of all cases are presented in Fig.~\ref{fig:biaxial-L}.
These curves evince that the phase-field model is not only able to simulate pressure-dependent peak and residual strengths, but also nearly insensitive to $L$ in every confining pressure case.
It is observed that compared with the previous example, the length parameter has a slightly greater influence on the softening response, although it does not affect the peak and residual strengths.
Such mild length sensitivity is natural and unavoidable for problems that deviate from the 1D homogeneous setting for which the specific forms of $g(d)$ and $H_{t}$ have been derived.
Note that similar levels of length sensitivity have also been observed for the existing phase-field models of cohesive opening fracture (\eg~see Fig. 16 of Geelen \etal~\cite{Geelen2019}).
Nevertheless, from a practical point of view, the phase-field model simulates pressure-dependent stress--strain--strength responses of brittle geomaterials without length sensitivity, allowing $L$ to be regarded as a purely geometric parameter.

\begin{figure}[htbp]
    \centering
    \includegraphics[width=1.0\textwidth]{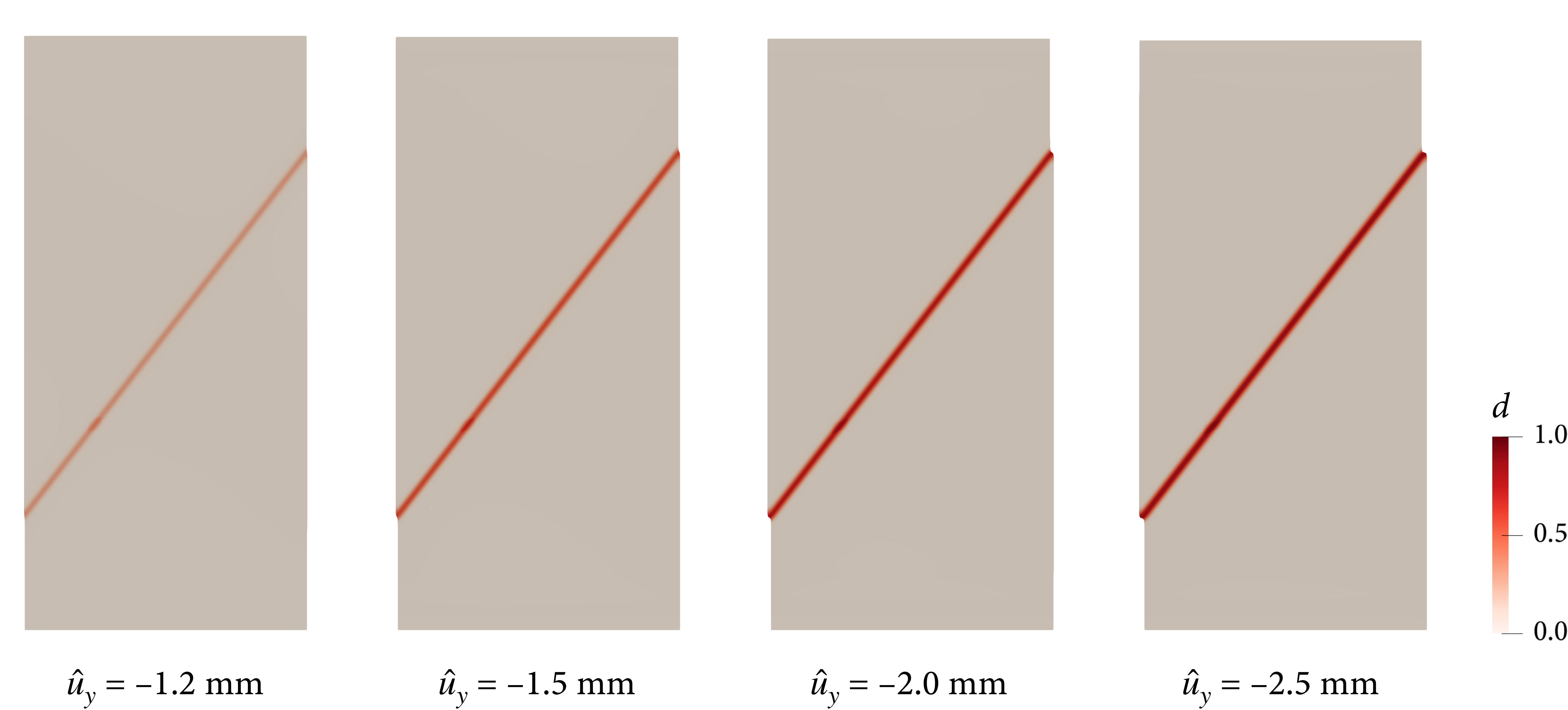}
    \caption{Biaxial compression: evolution of the phase field from $\hat{u}_{y} = -1.2$ mm to $\hat{u}_{y} = -2.5$ mm. $p_{c}=100$ kPa and $L=1$ mm. The domain is scaled by the displacement field.}
    \label{fig:biaxial-contours-pf-softening}
\end{figure}

\begin{figure}[htbp]
    \centering
    \includegraphics[width=1.0\textwidth]{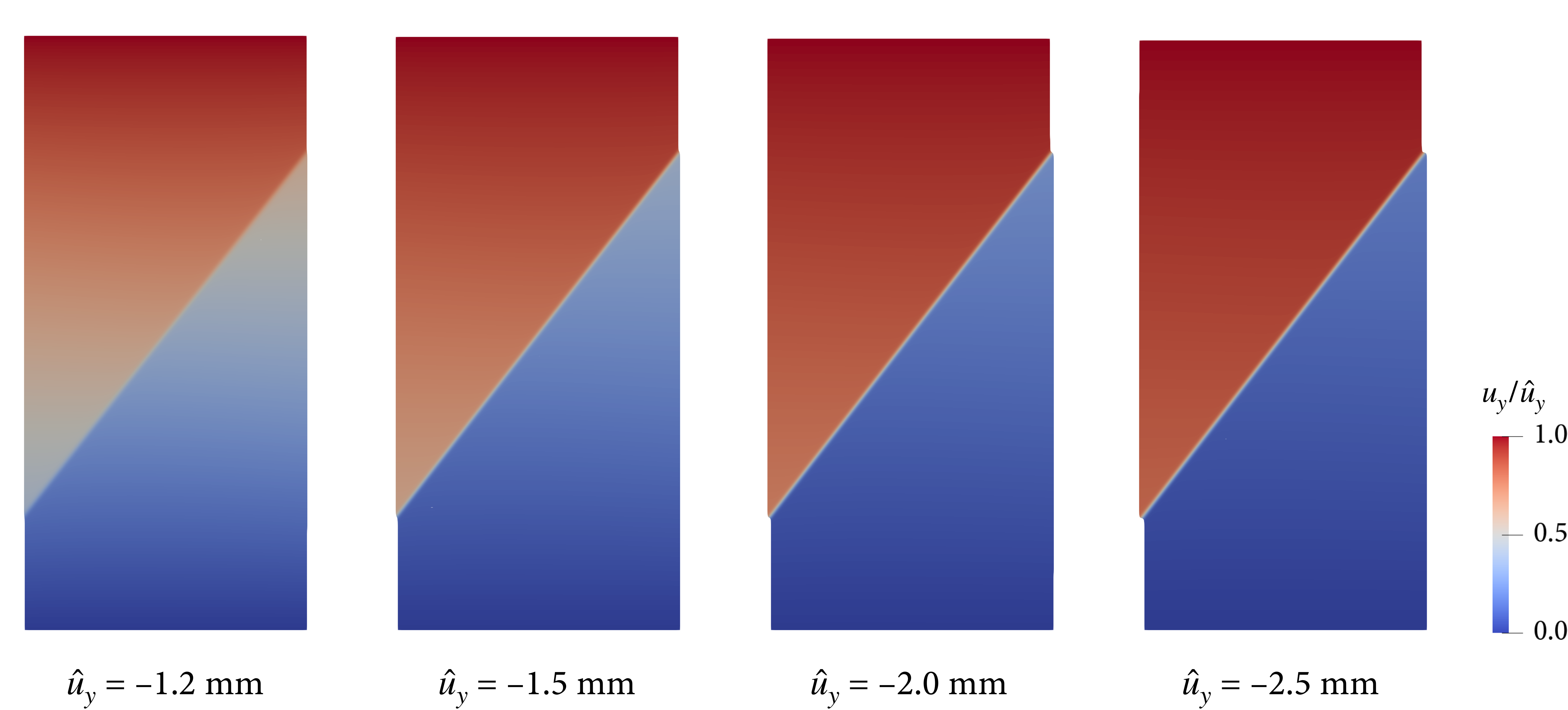}
    \caption{Biaxial compression: evolution of the normalized vertical displacement field from $\hat{u}_{y} = -1.2$ mm to $\hat{u}_{y} = -2.5$ mm. $p_{c}=100$ kPa and $L=1$ mm. The domain is scaled by the displacement field.}
    \label{fig:biaxial-contours-ydisp-softening}
\end{figure}

\begin{figure}[htbp]
    \centering
    \includegraphics[width=1.0\textwidth]{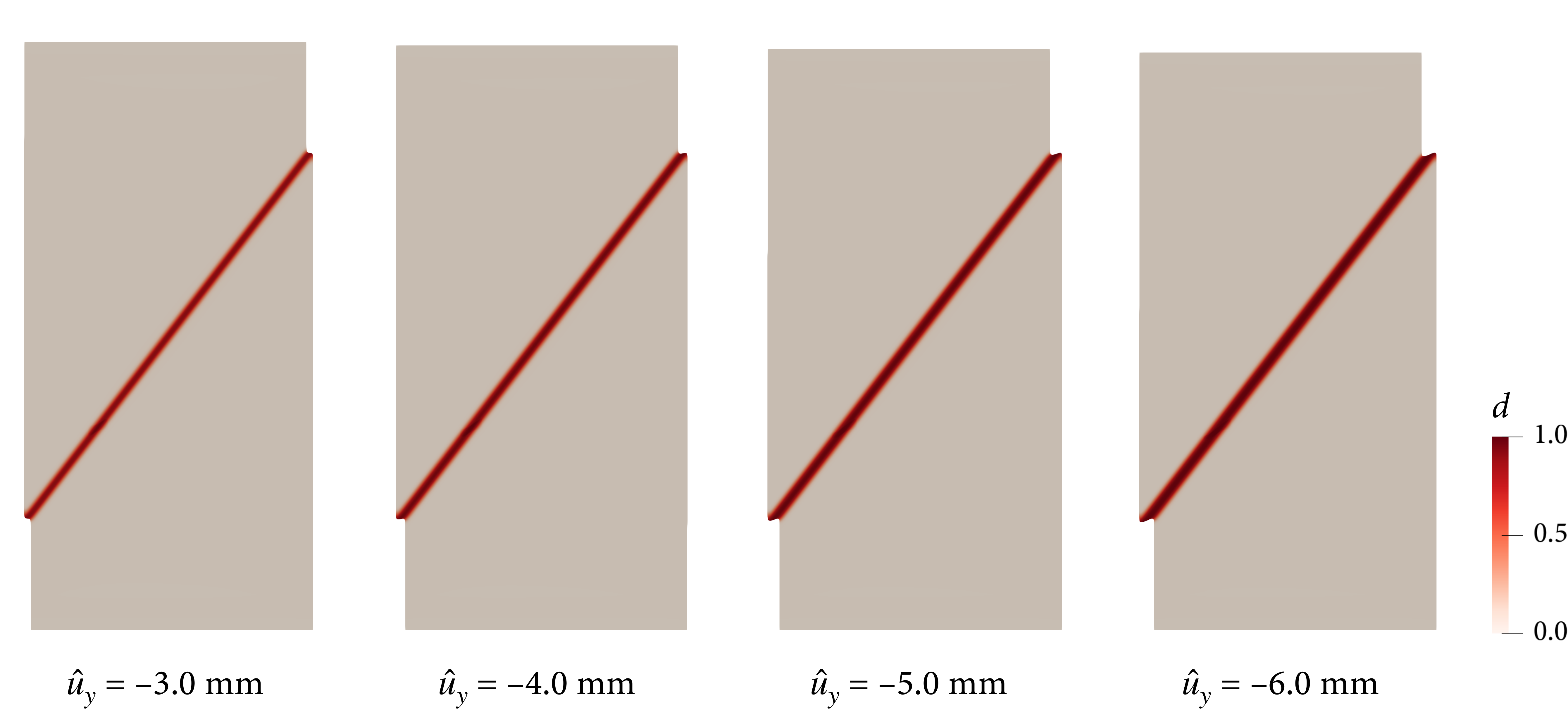}
    \caption{Biaxial compression: evolution of the phase field from $\hat{u}_{y} = -3.0$ mm to $\hat{u}_{y} = -6.0$ mm. $p_{c}=100$ kPa and $L=1$ mm. The domain is scaled by the displacement field.}
    \label{fig:biaxial-contours-pf-residual}
\end{figure}

\begin{figure}[htbp]
    \centering
    \includegraphics[width=1.0\textwidth]{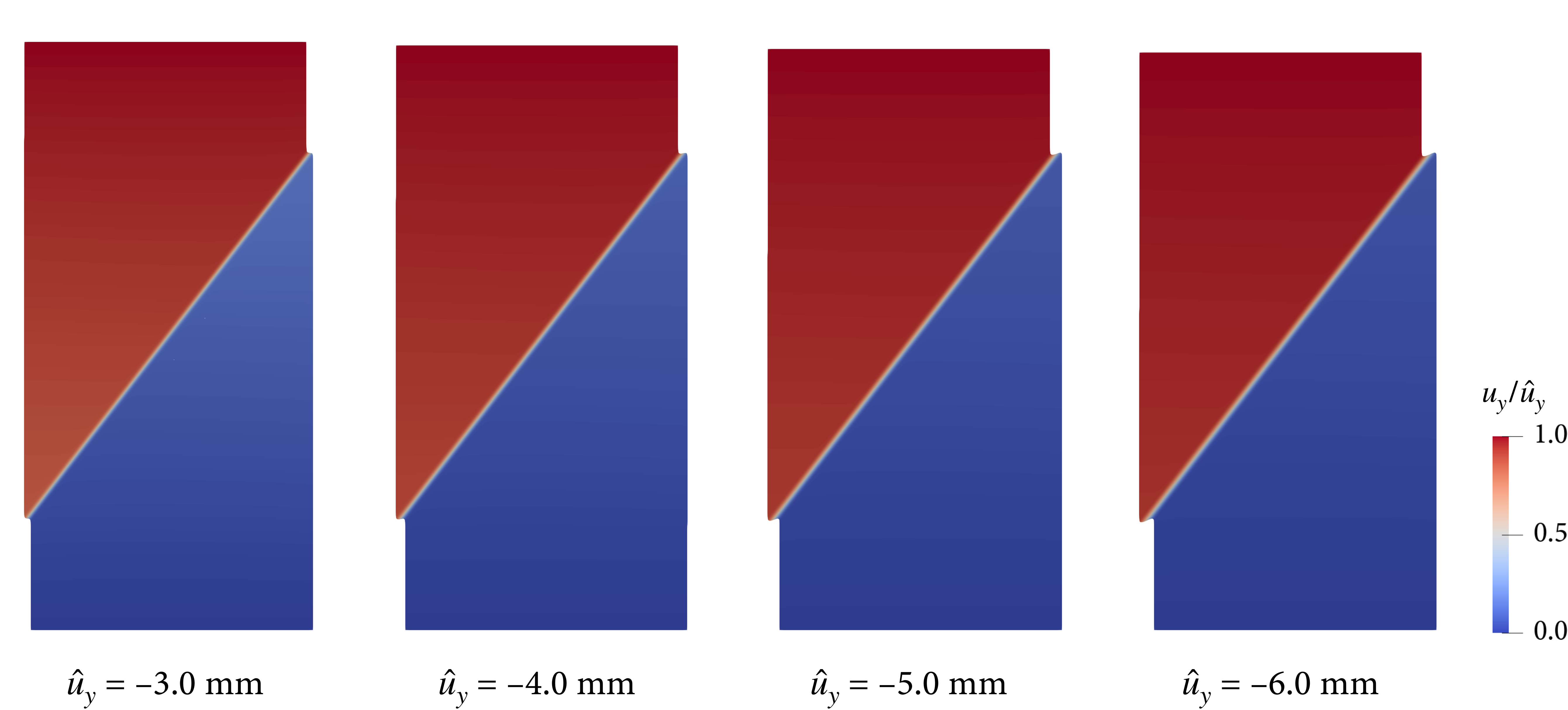}
    \caption{Biaxial compression: evolution of the normalized vertical displacement field from $\hat{u}_{y} = -3.0$ mm to $\hat{u}_{y} = -6.0$ mm. $p_{c}=100$ kPa and $L=1$ mm. The domain is scaled by the displacement field.}
    \label{fig:biaxial-contours-ydisp-residual}
\end{figure}

Further, to describe how the model simulates the shear fracturing process, we show in Figs.~\ref{fig:biaxial-contours-pf-softening}--\ref{fig:biaxial-contours-ydisp-residual} the evolutions of the phase field and the vertical displacement field when $p_{c}=100$ kPa and $L=1$ mm.
The focus of the first two figures (Figs.~\ref{fig:biaxial-contours-pf-softening} and~\ref{fig:biaxial-contours-ydisp-softening}) is on the main softening stage, and that of the next two figures (Figs.~\ref{fig:biaxial-contours-pf-residual} and~\ref{fig:biaxial-contours-ydisp-residual}) is on the subsequent stage in which the material asymptotically approaches its residual strength.
Figure~\ref{fig:biaxial-contours-pf-softening} shows that during the main softening stage, a process zone develops along the direction of $45^{\circ}-\phi_{r}/2$ degrees from the vertical, and the weak zone becomes fully fractured first.
The vertical displacement field in Fig.~\ref{fig:biaxial-contours-ydisp-softening} indicates that the upper and lower parts of the specimen are increasingly disconnected during this stage.
Once the specimen is almost fully fractured as in Fig.~\ref{fig:biaxial-contours-pf-residual}, the upper and lower blocks slide each other along the discontinuity as shown in Fig.~\ref{fig:biaxial-contours-ydisp-residual}.
Note that the sliding behavior is remarkably well simulated throughout the problem, which proves that the current model is built correctly on the stress-decomposition scheme for frictional contact~\cite{Fei2020}.
Other confining pressure cases show the essentially same behavior except that their peak and residual strengths are different.

Lastly, as can be seen from the force--displacement curves (Fig.~\ref{fig:biaxial-L}) as well as the phase-field evolution (Figs.~\ref{fig:biaxial-contours-pf-softening} and~\ref{fig:biaxial-contours-pf-residual}), the softening process is abrupt initially but prolonged until the stress reaches the residual stress.
This softening behavior emanates particularly from the chosen form of $g(d)$, namely the quasi-quadratic degradation function, which decays rapidly during the initial fracturing stage but slowly in the  (\cf~Fig.~\ref{fig:degradation-comparison-value}).
As such, the choice of $g(d)$ plays a critical role in the simulated material behavior, especially for larger-scale problems where the stress mobilized along the (potential) failure surface may vary more significantly.
This aspect is further investigated and demonstrated in the following example.

\subsection{Slope failure}
\begin{figure}[htbp]
    \centering
    \includegraphics[width=0.65\textwidth]{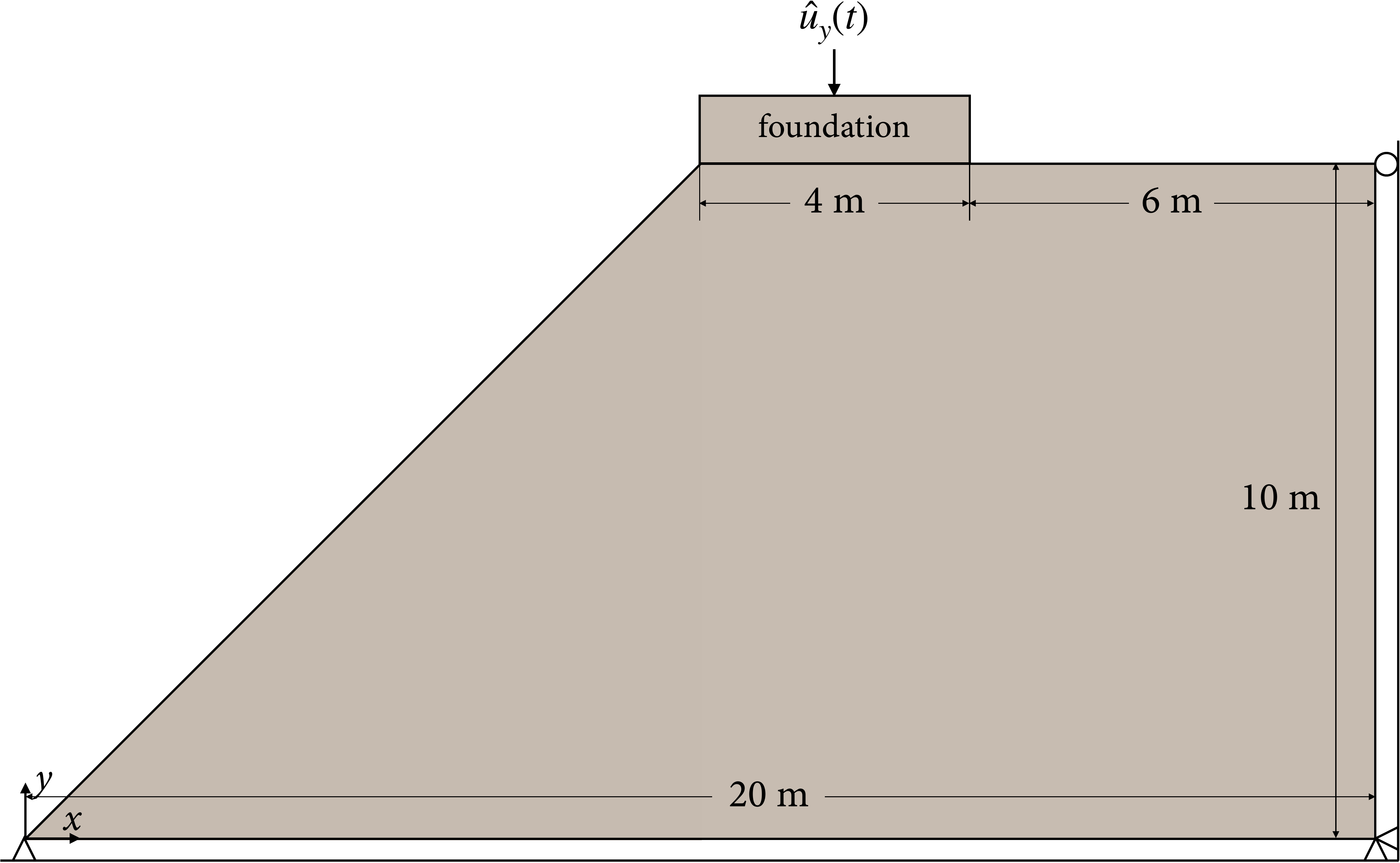}
    \caption{Slope failure: problem geometry and boundary conditions.}
    \label{fig:slope-setup}
\end{figure}

In our final example, we apply the phase-field model to the problem of slip surface growth at the field scale.
The purpose of this example is twofold: (i) to validate the capability of the phase-field model for simulating the propagation of a curved discontinuity passing through a domain with varied confining pressures,
and (ii) to investigate how the failure behavior is affected by the degradation function, $g(d)$.
For this purpose, we adapt the slope failure problem from Regueiro and Borja~\cite{Regueiro2001}, which was originally simulated by Drucker--Prager plasticity combined with bifurcation analysis and the strong discontinuity approach.
In this problem, a slope underlies a 4-m wide rigid foundation whose top middle point is pushed downward, see Fig.~\ref{fig:slope-setup}.
The geometry and boundary conditions give rise to a unique slip surface emerging from the right end of the slope--foundation interface.
The gravitational force is applied herein, making the confining pressure---and thus the peak and residual strengths---increasing with depth.

To investigate the effect of the degradation function $g(d)$ on the overall behavior, we simulate two cases, one with the quasi-quadratic $g(d)$ with $p=1$, and the other with the quasi-linear $g(d)$.
All other material and numerical parameters are the same in both cases, which are set as follows.
For material parameters shared with conventional elasto-plastic modeling, we adopt their values from Regueiro and Borja~\cite{Regueiro2001}: the cohesion strength $c=40$ kPa, the peak friction angle $\phi=16.7^{\circ}$, the residual friction angle $\phi_{r}=10^{\circ}$, the plane-strain Young's modulus $E=10$ MPa, Poisson's ratio $\nu=0.4$, and the mass density $\rho=2.04$ Mg/m$^3$.
These material parameters are similar to those of an overconsolidated clay.
However, the shear fracture energy $\sfe$ is not a parameter of Drucker--Prager plasticity.
Because $\sfe$ increases with the slip magnitude, we scale its value in the previous laboratory-scale examples to this field-scale problem.
Using an empirical relationship between the shear fracture energy and the slip magnitude in Abercrombie and Rice~\cite{Abercrombie2005}, we use $\sfe=10$ kJ/m$^2$.
For simplicity, we assume that $\sfe$ is constant although it may increase with normal stress within the same order of magnitude~\cite{Wong1986,Zhang1990}.
The phase-field length parameter is set to $L=0.2$ m, which is sufficiently lower than the upper bounds on $L$ for both the quasi-quadratic and quasi-linear $g(d)$.
We use a structured mesh, locally refining a region accommodating the expected slip surface to satisfy $L/h=10$.
The numerical results have been confirmed to show little sensitivity to $L$ and $h$, as long as their values are reasonable.
The simulation is conducted applying a constant displacement rate of $4\times10^{-4}$ m to the top middle of the foundation until the displacement reaches 0.3 m.
Note that the subsequent process, which involves rapid sliding of a failed soil mass, may not be well modeled by the current formulation which is restricted to infinitesimal strain, quasi-static conditions.

\begin{figure}[htbp]
    \centering
    \includegraphics[width=0.55\textwidth]{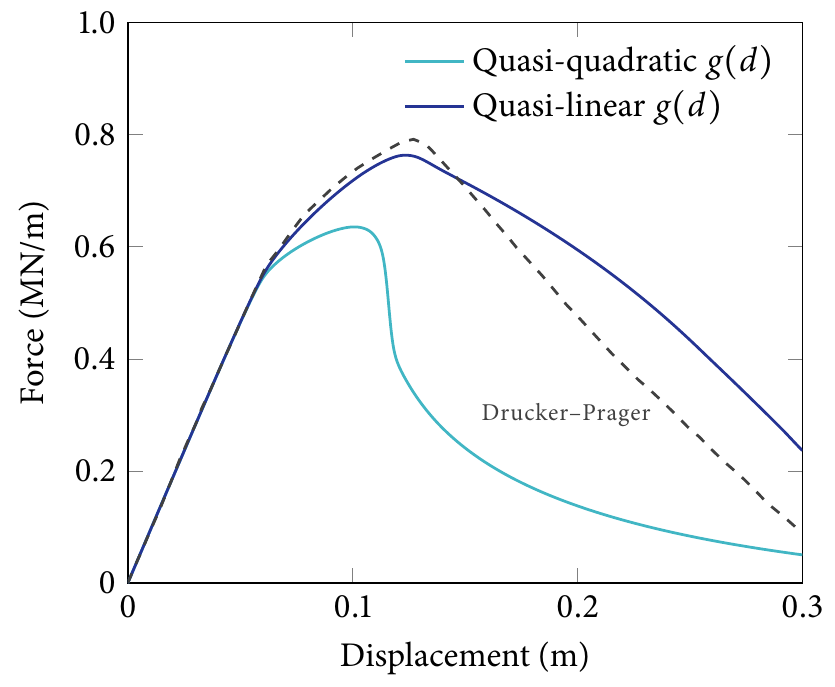}
    \caption{Slope failure: force--displacement curves produced with the quasi-quadratic and quasi-linear degradation functions. The dashed line denotes the result obtained by Regueiro and Borja~\cite{Regueiro2001} using Drucker--Prager plasticity and the strong discontinuity approach.}
    \label{fig:slope-force-disp}
\end{figure}

Figure~\ref{fig:slope-force-disp} compares the force--displacement curves obtained with the quasi-quadratic and quasi-linear degradation functions.
Also presented here is the curve digitized from Regueiro and Borja~\cite{Regueiro2001} whereby Drucker--Prager plasticity is used in conjunction with the strong discontinuity approach.
It is found that, despite the same material parameters, the quasi-linear $g(d)$ renders the overall structural behavior significantly less brittle.
This trend is consistent with that the quasi-linear $g(d)$ decays more slowly than the quasi-quadratic $g(d)$ (\cf~Fig.~\ref{fig:degradation-comparison-value}), as well as that the quasi-linear $g(d)$ leads to a higher peak load for opening fracture (see Fig. 18 of Geelen \etal~\cite{Geelen2019}, for example).
Interestingly, the pre-peak behavior of the quasi-linear result is very similar to that of the plasticity result, even though the two modeling approaches rely on completely different criteria for the initiation and propagation of shear discontinuities.
It is also observed that the quasi-quadratic $g(d)$ leads to a long asymptotic behavior, as in the previous examples.

\begin{figure}[htbp]
    \centering
    \includegraphics[width=0.8\textwidth]{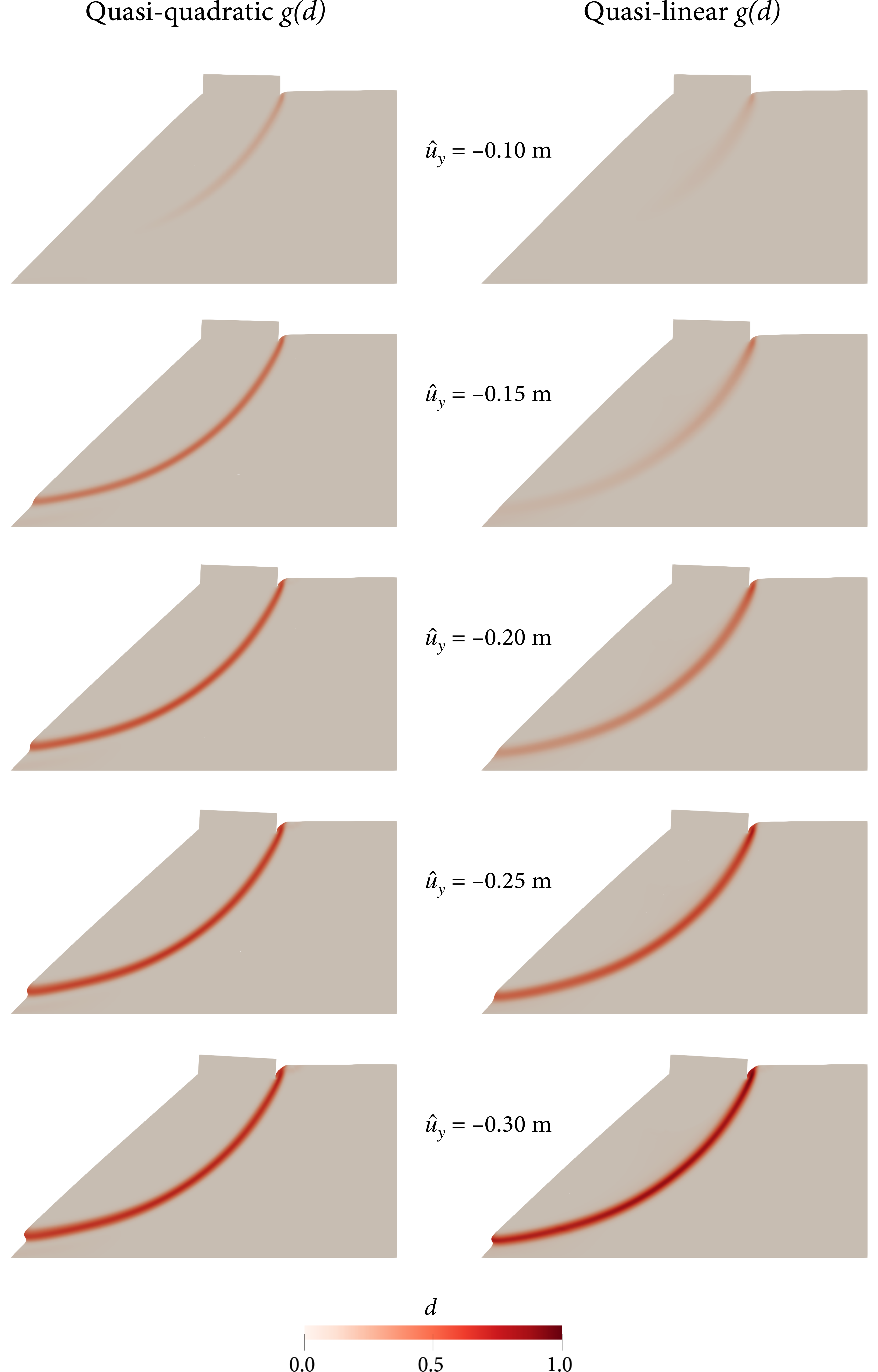}
    \caption{Slope failure: evolution of the phase field. The domain is scaled by the displacement field with a factor of 2.}
    \label{fig:slope-contours-pf}
\end{figure}

\begin{figure}[htbp]
    \centering
    \includegraphics[width=0.8\textwidth]{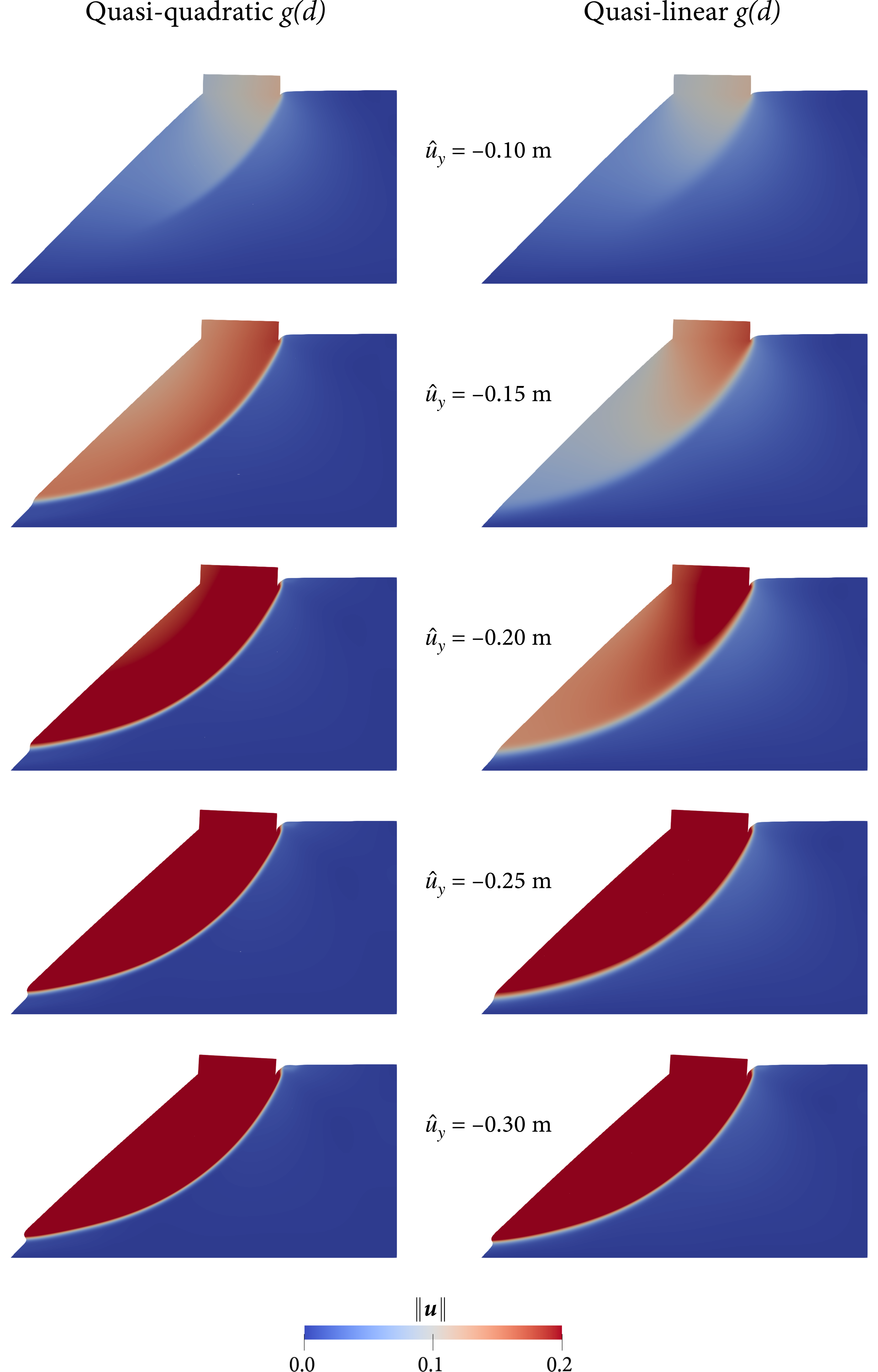}
    \caption{Slope failure: evolution of the displacement magnitude. The domain is scaled by the displacement field with a factor of 2. Color bar in meter.}
    \label{fig:slope-contours-disp}
\end{figure}

Figures~\ref{fig:slope-contours-pf} and~\ref{fig:slope-contours-disp} show how the phase field and the displacement magnitude evolve during the failure process.
These figures indicate that the nonlinear part of the pre-peak behavior corresponds to the stage in which the slip process zone emerges from the slope--foundation interface and propagates to the left boundary of the slope ($\hat{u}_{y}=-0.10$ m in the figures).
One can also see that the process zone propagates faster when the quasi-quadratic $g(d)$ is used, which is consistent with the more brittle behavior of the quasi-quadratic $g(d)$ in Fig.~\ref{fig:slope-force-disp}.
Once the slip process zone reaches the slope boundary, it gives rise to post-peak softening behavior of the overall slope system, and develops further into a fully-discontinuous slip surface.
Notably, although the process zone in the quasi-linear $g(d)$ result propagates more slowly, it grows more rapidly into a full slip surface, which can also be gleaned from the softening rate of the force--displacement curves in Fig.~\ref{fig:slope-force-disp}.
This tendency is also consistent with the difference between the variations of the quasi-quadratic and quasi-linear $g(d)$ with $d$, depicted in Fig.~\ref{fig:degradation-comparison-value}.
Also importantly, Fig.~\ref{fig:slope-contours-disp} demonstrates that the phase-field model well simulates the sliding behavior of the failed soil mass along the curved slip surface.
This capability is a remarkable feature in that no algorithm was necessary for such robust simulation of frictional sliding.

\begin{figure}[htbp]
    \centering
    \includegraphics[width=1.0\textwidth]{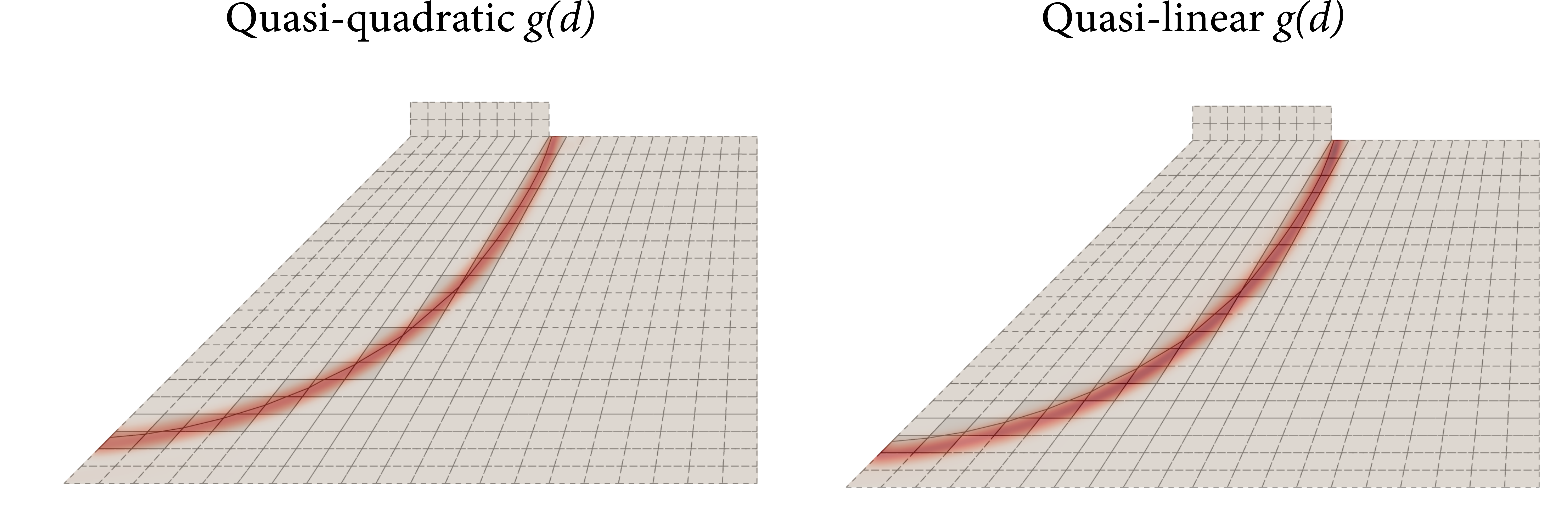}
    \caption{Slope failure: comparison of the slip surfaces from the quasi-quadratic and quasi-linear degradation functions. The results are superimposed on the mesh in Regueiro and Borja~\cite{Regueiro2001} where bifurcated/enhanced elements are shaded.}
    \label{fig:slope-comparison}
\end{figure}

In Fig.~\ref{fig:slope-comparison}, the slip surfaces obtained by the two degradation functions are compared using the phase-field values at the displacement of 0.3 m.
As a reference, the mesh in Regueiro and Borja~\cite{Regueiro2001}, in which bifurcated/enhanced elements are shaded to denote the slip surface in their simulation, is laid over the phase-field values.
It is found that both slip surfaces compare remarkably well with the bifurcation path of the Drucker--Prager plasticity model.
This agreement indicates that the fracture path of the phase-field model is generally consistent with bifurcation theory.
It can also be seen that around the final path (lower left part), the slip surface of the quasi-linear $g(d)$ result is located slightly below than that of the quasi-quadratic $g(d)$.
Given that the slip surface has propagated more slowly with the quasi-linear $g(d)$, this difference agrees with the fact that the localized failure becomes more deep-seated as the propagation becomes more delayed~\cite{Petley1997,Elbedoui2009}.
This consistency further validates the proposed phase-field model.

Three important conclusions can be drawn from this example.
First, the phase-field model can well simulate the initiation and propagation of a slip surface at the field scale, without any algorithm for geometry tracking or frictional contact, relying on the fracture mechanics theory of Palmer and Rice~\cite{Palmer1973}.
Second, the softening behavior at the material level, which is characterized by $g(d)$ in the phase-field model, has marked impacts on the strength and brittleness of an overall structural system, because it determines the stress mobilized along a (potential) slip surface.
Third, the fracture mechanics and plasticity/bifurcation approaches can lead to significantly different results, even when the locations of the slip surfaces are virtually identical and the pre-peak behaviors are quite similar.
Together with the second conclusion, the difference highlights the critical role of the propagation dynamics and frictional sliding in geologic shear fracture processes.
Although the fracture mechanics approach arguably lends itself to more physically-based analysis of the propagation and sliding processes, extensive research will be required to validate and comparatively evaluate the two approaches for modeling the initiation and growth of shear discontinuities at the field scale.

\section{Closure}
\label{sec:closure}
A new phase-field model has been developed for numerical simulation of geologic shear fractures such as faults and slip surfaces.
The new model has three features that are critical to shear fractures in geologic materials but absent in previous phase-field models of fracture.
They are: (i) friction along crack surfaces and its impact on the fracture propagation mechanism, (ii) insensitivity to the phase-field length parameter, and (iii) pressure dependence of the peak and residual shear strengths.
As demonstrated by numerical examples, the combination of these three features renders the proposed model a theoretically rigorous and computationally efficient method to simulate shear fracture processes in geologic materials.

Future work will focus on advancing the model to accommodate inertial effects and coupling with fluid flow in porous media.
Incorporation of these two additional features will enable the phase-field model to be applicable to a wider variety of geologic hazards problems from progressive failure in low-permeable geomaterials to injection-induced earthquakes.
Work is underway to extend the current model to the dynamic regime as well as to combine it with robust and efficient poromechanical formulations at small and large strains~\cite{Choo2018c,Choo2018d,Choo2019,Zhao2020}.
Successful results will be published in future publications.

\section*{Acknowledgments}
The authors are grateful to Professor Teng-fong Wong for motivating this study and sharing his insight.
Thanks also go to graduate student Yidong Zhao for proofreading the manuscript.
This work was supported by the Research Grants Council of Hong Kong (Projects 27205918 and 17201419).
The first author also acknowledges financial support from his Hong Kong PhD Fellowship.

\section*{Appendix. Analysis of the phase-field model in 1D simple shear}
\label{sec:appendix}
This section provides mathematical analyses of the proposed phase-field model of frictional shear fracture in terms of fracture energy dissipation and slip displacement.
Drawing heavily on the methodology of Geelen \etal~\cite{Geelen2019} for 1D analysis of their phase-field model of cohesive opening fracture, we analyze our own model of shear fracture in a 1D simple shear setting.

\subsection*{Fracture energy dissipation}
To find an expression for the energy dissipated during fracturing in our phase-field model, we recall the equations and notations of the 1D simple shear problem discussed in Section~\ref{subsec:derivation-gd-Ht} and the conclusion of $\Lambda=0$.
At a material point where shear fracture is fully formed, \ie~$d^{*} = 1$, the shear stress therein is equal to the residual shear strength, \ie~$\tau=\tau_r$.
In this case, Eq. \eqref{eq:micro-force-1d-v2} reduces to
\begin{align}
    d_u(x) = L^2\left(\dfrac{\od d_u(x)}{\od x} \right)^2\,,
    \label{eq:ultimate-damage-distribution}
\end{align}
where $d_u(x)$ denotes the ultimate distribution of $d$ along the interface normal direction.
The energy dissipated by fracturing along the $x$-direction is calculated as
\begin{align}
    \mathcal{D}_{\text{frac}} &= \int_\Omega \sfe\gamma(d,\, \grad{d})\, \od V
    = \int^{b}_{-b} \dfrac{3\sfe}{8} \left[\dfrac{d_u(x)}{L} + L \left(\dfrac{\od d_u(x)}{\od x} \right)^2\right] \, \od x\,. \label{eq:fracture-dissipation}
\end{align}
Then, by inserting Eq. \eqref{eq:ultimate-damage-distribution} into Eq. \eqref{eq:fracture-dissipation} and invoking the symmetry of $d_u(x)$, we get
\begin{align}
    \mathcal{D}_{\text{frac}}  &= \int^{b}_{-b} \dfrac{3\sfe}{8} \left[\dfrac{d_u(x)}{L} + L \left(\dfrac{\od d_u(x)}{\od x} \right)^2\right] \, \od x \nonumber\\
    &= \int_0^b \dfrac{3\sfe}{4} \left[\dfrac{d_u(x)}{L} + L \left(\dfrac{\od d_u(x)}{\od x} \right)^2\right] \, \od x \nonumber\\
    &= \dfrac{3\sfe}{4}\int_0^b 2\sqrt{d_u(x)} \dfrac{\od d_u(x)}{\od x} \, \od x \nonumber\\
    &= \dfrac{3\sfe}{2}\int_0^1 \sqrt{d_u} \, \od d_u \nonumber\\
    &= \sfe\,.
\end{align}
This proves that the fracture energy dissipation equals the shear fracture energy, $\sfe$, as prescribed in the model.
This equivalence has also been verified numerically through the first example in Section~\ref{sec:examples}.

\subsection*{Slip displacement}
We further derive the apparent slip displacement of the phase-field model based on the 1D simple shear problem.
To this end, let us first consider the total displacement $u^*$ at the free end of the 1D domain undergoing simple shear.
When the maximum value of $d$ is $d^*$, we can calculate $u^*$ by integrating the shear strain over the height of the domain ($h$), \ie~
\begin{align}
    u^* &= \int^{h/2}_{-h/2} \gamma \, \od x = 2 \int^{h/2}_0 \gamma \, \od x\,.
    \label{eq:total-u-1d-v1}
\end{align}
Recalling $\tau_\bulk = G\gamma$ as well as Eq. \eqref{eq:tau-bulk-1d}, $\gamma$ can be expressed as
\begin{align}
    \gamma = \dfrac{\tau - \tau_r}{Gg(d)} + \dfrac{\tau_r}{G} \, .
\end{align}
Inserting the above equation into Eq.~\eqref{eq:total-u-1d-v1}, $u^*$ can be written as
\begin{align}
    u^* &= \dfrac{2}{G} \int^{h/2}_0 \left(\dfrac{\tau(d^*) - \tau_r}{g(d)} + \tau_r \right) \, \od x
    \label{eq:total-u-1d-v2} \,,
\end{align}
with an assumption that $G$ is constant across the height of the domain.
From the degradation function~\eqref{eq:degradation-function-form}, we have
\begin{align}
    \frac{1}{g(d)} = 1 + \phi(d) \,, \quad \text{where}\;\; \phi(d) := \dfrac{md(1+pd)}{(1-d)^n}\, .
\end{align}
Substituting this expression into Eq. \eqref{eq:total-u-1d-v2} gives
\begin{align}
    u^* = {h\dfrac{\tau(d^*)}{G}} +  \underbrace{\dfrac{2}{G}\int^{h/2}_0\phi(d)(\tau(d^*) - \tau_r) \, \od x}_{=:\delta}.
\end{align}
In the above equation, the first term on the right hand side corresponds to the continuous shear deformation, whereas the second term to the discontinuous slip displacement, which will be denoted by $\delta$ in the following.
By changing the integration variable from $\od x$ to $\od d$, we can write $\delta$ alternatively as
\begin{align}
    \delta &= \dfrac{2[\tau(d^*) - \tau_r]}{G}\int^{h/2}_0 \phi(d) \dfrac{\od x}{\od d} \, \od d \, . \label{eq:slip-dd}
\end{align}
Combining Eqs.~\eqref{eq:tau-derived-v1} and~\eqref{eq:micro-force-1d-v2} and noting that $\Lambda = 0$, we obtain
\begin{align}
    \dfrac{\od d}{\od x} = - M(d^*,\, d)^{1/2}\,, \quad \text{where}\;\;
    M(d^*,\, d) := \dfrac{1}{L^2}\left[d - \dfrac{g(d)^{-1} - 1}{g(d^*)^{-1} - 1}d^*\right]\,.
    \label{eq:d-derivatives}
\end{align}
Substituting Eq. \eqref{eq:d-derivatives} into Eq. \eqref{eq:slip-dd} gives
\begin{align}
    \delta = \dfrac{2[\tau(d^*) - \tau_r]}{G}\int^{d^*}_0 \phi(d) M(d^*,\, d)^{-1/2} \, \od d \, . \label{eq:slip-dd-tau}
\end{align}
We then insert $\tau(d^\ast)$ in Eq. \eqref{eq:tau-derived-v3} into Eq. \eqref{eq:slip-dd-tau} and arrive at
\begin{align}
    \delta &= \dfrac{2(\tau_p - \tau_r)Lm}{G}\int^{d^\ast}_{0} \sqrt{\dfrac{d^\ast}{d\phi(d^\ast) - d^\ast\phi(d)}}\phi(d) \, \od d \\
    &= \dfrac{\sfe}{\tau_p - \tau_r}\frac{3}{2}\sqrt{\dfrac{1}{m}} \int^{d^\ast}_{0} \sqrt{\dfrac{d^\ast}{d\phi(d^\ast) - d^\ast\phi(d)}}\phi(d) \, \od d \, . \label{eq:slip-dd-taup}
\end{align}
Remarkably, the coefficient of the final expression, $\sfe/(\tau_p - \tau_r)$, is equal to the characteristic slip displacement derived in Palmer and Rice~\cite{Palmer1973} (denoted by $\bar{\delta}$ therein).

\bibliography{references}

\end{document}